%% file: tamura201811arx.tex
\documentclass[]{pasj01}  
\draft 

\pdfoutput=1

\newcommand{\lgauss}{$\sigma_{\rm Line}$ }
\newcommand{\sigL}{$\sigma_{\rm Line}$ }

\newcommand{\deltaEnar}{$\Delta E_{\rm narrow}$ }
\newcommand{\dEW}{$\Delta {\rm EW}$ } 
\newcommand{\Eline}{$E_{\rm Line}$ }
\newcommand{\EW}{$EW$ } 

\newcommand{\suzaku}{{Suzaku }}

\newcommand{\hitomi}{{Hitomi }}
\newcommand{\xarm}{{XRISM }}
\newcommand{\atomdb}{\texttt{AtomDB} }
\newcommand{\spex}{\texttt{SPEX} }
\newcommand{\kms}{{\rm km s$^{-1}$} }

\newcommand{\dmDecay}{$\Gamma _{\rm DM}$\ }
\newcommand\jcap{Journal of Cosmology Astroparticle Physics}%
\newcommand{\UNITPFLUX}{{\rm counts cm$^{-2}$ s$^{-1}$}}

\begin{document}
\SetRunningHead{Author(s) in page-head}{Running Head}

\title{
  An X-ray spectroscopic search for dark matter and
  unidentified line signatures in the Perseus cluster with \hitomi}

\author{
  Takayuki \textsc{Tamura}\altaffilmark{1}, 
  Andrew C. \textsc{Fabian}\altaffilmark{8},
  Poshak \textsc{Gandhi}\altaffilmark{2}, 
  Liyi \textsc{Gu}\altaffilmark{3},
  Ayuki \textsc{Kamada} \altaffilmark{4},
  Tetsu \textsc{Kitayama}\altaffilmark{5},
  Michael \textsc{Loewenstein}\altaffilmark{6,11},
  Yoshitomo \textsc{Maeda}\altaffilmark{1}, 
  Kyoko \textsc{Matsushita}\altaffilmark{7},
  Dan \textsc{McCammon}\altaffilmark{10}, 
  Kazuhisa \textsc{Mitsuda}\altaffilmark{1},
  Shinya \textsc{Nakashima}\altaffilmark{3},
  Scott \textsc{Porter}\altaffilmark{6}, 
  Ciro \textsc{Pinto}\altaffilmark{8},
  Kosuke \textsc{Sato}\altaffilmark{9},
  Francesco \textsc{Tombesi}\altaffilmark{6},
  and
  Noriko \textsc{Y. Yamasaki}\altaffilmark{1}
}
\altaffiltext{1}
{
Institute of Space and Astronautical Science,
Japan Aerospace Exploration Agency,\\
3-1-1 Yoshinodai, Chuo-ku, Sagamihara, Kanagawa 229-8510, Japan
}
\email{tamura.takayuki@jaxa.jp}
  \altaffiltext{2}{
Department of Physics \& Astronomy, 
University of Southampton, 
SO17 1BJ, UK
}
\altaffiltext{3}{RIKEN High Energy Astrophysics Laboratory, 2-1 Hirosawa, Wako, Saitama 351-0198, Japan}
  \altaffiltext{4}{Center for Theoretical Physics of the Universe, Institute for Basic Science (IBS), Daejeon 34126, Korea}
  \altaffiltext{5}{Department of Physics, Toho University,  2-2-1 Miyama, Funabashi, Chiba 274-8510}
  \altaffiltext{6}{NASA, Goddard Space Flight Center, 8800 Greenbelt Road, Greenbelt, MD 20771, USA}
  \altaffiltext{7}{Department of Physics, Tokyo University of Science, 1-3 Kagurazaka, Shinjuku-ku, Tokyo 162-8601}
  \altaffiltext{8}{Institute of Astronomy, Madingley Road, CB3 0HA Cambridge, United Kingdom}
  \altaffiltext{9}{Department of Physics, Saitama University, 255 Shimo-Okubo, Sakura-ku, Saitama, 338-8570}
  \altaffiltext{10}{Physics Department, University of Wisconsin, Madison, WI 53706, USAS}
  \altaffiltext{11}{University of Maryland College Park, Department of Astronomy, 4296
Stadium Dr., College Park, MD , USA, 20742}

\maketitle

\begin{abstract}
We present results of a search for unidentified line emission and
absorption signals in the 2--12~keV energy band of spectra extracted
from Perseus Cluster core region observations obtained with the 5~eV
energy resolution \hitomi Soft X-ray Spectrometer.
No significant
unidentified line emission or absorption is found.
Line flux upper limits (1$\sigma$ per resolution element)
vary with photon energy and assumed intrinsic width, 
decreasing from $\sim 100$ photons cm$^{-2}$ s$^{-1}$ sr $^{-1}$ at 2~keV to $<10$
cm$^{-2}$ s$^{-1}$ sr $^{-1}$ over most of the 5--10 keV energy range
for a Gaussian line with Doppler broadening of 640 \kms.
Limits for
narrower and broader lines have a similar energy dependence
and are systematically smaller and larger, respectively.
These line flux limits are used to constrain
the decay rate of hypothetical dark matter candidates.
For the sterile neutrino decay rate,
new constraints over the the mass range of 4--24~keV with mass resolution better than any previous X-ray analysis
are obtained.
Additionally, 
the accuracy of relevant thermal spectral models and atomic data are evaluated.
The Perseus cluster spectra may be described by a composite of multi-temperature thermal and AGN
power-law continua.
Superposed on these, a few line emission signals possibly originating
from unmodeled atomic processes 
(including Si \emissiontype{XIV} and Fe \emissiontype{XXV})
 are marginally detected and tabulated.
Comparisons with previous X-ray upper limits and
future prospects for dark matter searches using high-energy resolution spectroscopy are discussed.
\end{abstract}
\section{Introduction}

X-ray spectroscopy
has long been used to study the physical state and
elemental composition in various astronomical hot plasmas.
The advent of high energy resolution instruments in orbit 
has improved the sensitivity
to weak spectral features, 
and hence the diagnostic power of X-ray spectroscopy.
Most recently,
the Soft X-ray Spectrometer (SXS; \cite{Kelley2018}), 
the first X-ray calorimeter 
that collected observations
in orbit,
onboard the \hitomi observatory
\citep{Takahashi2018} 
demonstrated new capabilities by 
achieving a 5~eV FWHM energy resolution.
This represents
superior sensitivity and better resolving power above 2~keV than any previous X-ray grating instruments
and, by virtue of being non-dispersive, 
is better-suited for observing extended sources such as clusters of galaxies.

Thanks to 
these capabilities, \hitomi high-energy resolution
spectroscopic analysis
may be used to enhance dark matter X-ray searches.
There are
several reports of putative detections
of unidentified X-ray signals, 
such as \citet{2010ApJ...725L.131P}.
A recent, much discussed claim is of
a line at around 3.5~keV 
in the Perseus cluster and a stacked sample of clusters
by \citet{2014ApJ...789...13B}.
Together with subsequent reports
(e.g. \cite{2014PhRvL.113y1301B};
\cite{2015PhRvL.115p1301B}; \cite{2016PhRvD..94l3504N}
claiming independent detections
consistent with the dark matter interpretation,
these works have 
attracted a substantial number of follow-up observational studies
with different instruments and objects.
Among these are works by
  \citet{tamura2015}, \citet{sekiya2015}, and others
reporting non-detection and
disfavoring a substantial part of the claimed 3.5~keV flux range
(see \cite{2017PhR...711....1A} for summaries).
As discussed in 
these studies,
potential dark matter X-ray signals are expected to be
too faint 
to be resolved by previous X-ray detectors (mostly Charge-Coupled Devices; CCDs).
In fact, the claimed 3.5~keV signals
have line equivalent widths
of about 1~eV,
corresponding to a
$<1$\% excesses above continuum,
generally smaller than the instrumental calibration uncertainties
(typically $10$\%).
Even the deepest and targeted searches
such as \citet{2016MNRAS.460.1390R} and  \citet{2018ApJ...854..179C}
may have marginally detected a signal at 3.5~keV 
but failed to disentangle various possibilities
due to limited CCD energy resolution.

The unidentified X-ray line signals may originate from dark matter radiative decay,
as originally proposed by \citet{2001ApJ...562..593A}, 
with the keV sterile neutrino being one long-standing candidate 
(see \cite{2017JCAP...01..025A} for a
review).
The 3.5~keV claims
stimulate theoretical consideration of sterile neutrino production mechanisms, 
as well as many other dark matter candidate particles, 
conceived in order to explain unidentified X-ray signals.
These in turn
indicate
how parameter space can be constrained or new physics searches expanded
exclusively by X-ray spectroscopy of cosmic dark matter systems.
Many, but by no means all, 
these dark matter candidates tend to behave as warm dark matter,
suppressing sub-galactic structure formation by free-streaming.
Probes of small-scale matter clustering such as a number count of
satellite galaxies (e.g. \cite{2017PhRvD..95h3015C})
and Lyman-$\alpha$ forest measurements
(e.g., \cite{2017JCAP...12..013B} and references therein)
play a complementary role in searching for proposed dark matter candidates
(e.g., \cite{2018JCAP...01..054B} for further discussion).

In addition to enhancing the quality of X-ray searches for dark
matter, 
improvements of spectroscopic sensitivity 
provide an opportunity
to measure faint signals from relatively rare elements 
and previously unresolved atomic features \citep{2017Natur.551..478H}.
For example,  \citet{2014ApJ...789...13B}
discussed possible 3.5~keV line origins 
in atomic features from K, Ar, or Cl ions; 
while, 
\citet{2015A&A...584L..11G}  proposed an origin from 
charge exchange emission.
  Disentangling such mixtures of processes from these 
  dominant thermal emission
is only accessible via
high resolution spectroscopy,
as discussed thoroughly in 
\citet{2008SSRv..134..155K} and
\citet{2014arXiv1412.1172S}.

There are now a few calorimeter data sets of relevance to searches for dark
matter and other weak features.
\citet{2015ApJ...814...82F}
used sounding rocket calorimeter spectra
with energy resolution of 11--23~eV,
covering a wide sky area, 
with a short exposure time of 106~s
towards an anti-center Galactic plane region.
They revealed no 3.5~keV signal.
\citet{2017ApJ...837L..15A} 
used the SXS Perseus spectra
to search for the 3.5~keV signal, 
deriving an upper limit inconsistent
with previous claimed detections in Perseus.
Utilizing the latest versions of ATOMDB and SPEX plasma codes, 
\citet{H2018-T} (hereafter H2018-T) 
found that the same Perseus spectra may be described
by simple thermal models with an average temperature of 4~keV
without any significant residuals in the 2--15~keV  energy band.
Furthermore,
\citet{H2018-A} (hereafter H2018-A) 
evaluated spectral fit residuals
and the accuracy of the model input atomic physics.
\citet{2018-crab}
examined the SXS spectra of the Crab Nebula,
a spectrally featureless calibration source, 
searching for previously undetected  line emission or absorption signatures.
We use this result to estimate the calibration uncertainty
bf of the instrumental effective area.

The highest quality spectrum
with the \hitomi SXS
is of the core of Perseus, 
a nearby relaxed galaxy cluster 
and the X-ray brightest extra-galactic extended source \citep{h16nat}.
Making use of the extensive plasma modeling
conducted in \hitomi papers cited above, 
we further 
examine the spectra 
for weak line emission and absorption signals
at the energies of both known and unknown features.
We present the first line search results
over a broad X-ray energy band (2--12~keV)
from a massive dark matter structure 
at energy resolution better than any previous X-ray analysis.
We resolve a number of strong and weak atomic line emission features
 but  find no significant unidentified line emission nor absorption.
We derive upper limits used to constrain
the sterile neutrino decay rate and compare our results with previous X-ray limits.
 Discussion is made 
on the limitations of a weak line
search with an instrument with high energy resolution but modest collecting area.

Throughout this paper,
we assume the following cosmological parameters:
$H_0 = 70$ km s$^{-1}$Mpc$^{-1}$, 
$\Omega_\mathrm{m} = 0.3$, and $\Omega_\mathrm{\Lambda} = 0.7$.
One arc-minute corresponds to 21.1~kpc
at the Perseus cluster redshift, z = 0.017284 \citep{H2018-V}.
Unless otherwise stated, 
we use the 68\% ($1\sigma$) confidence level for errors
and single-line signal detection significance levels 
without accounting for the number of independent energy trial bins
(i.e., the look elsewhere effect)
,
and express X-ray energies in the observed (redshifted) frame.

\section{Observations and data analysis}

\subsection{Observations and spectral extraction}
 Th Perseus cluster was observed 
four times with the \hitomi SXS in 2016 February and March.
These observations are used
in the series of papers described in section 1.
Detailed descriptions of the \hitomi observatory and the SXS instrument
are are found in \citet{Takahashi2018} and \citet{Kelley2018}, respectively.
Hitomi's X-ray telescope references include \citet{2016SPIE.9905E..0ZO} and \citet{2018PASJ...70...19M}.

In H2018-T 
we processed and extracted  the relevant 
spectral and responses files
 which are used herein.
We describe the method briefly below.
To process data and response files, 
we use the HEAsoft version 6.21,
\hitomi software version 6,
and calibration database version 7 \citep{2018JATIS...4a1207A}.

We use Observations 
2 (obsid 100040020), 
3 (100040030, 100040040, 100040050),
and
4 (100040060) with 
a total exposure time of 289~ks.
These observations cover the cluster central core 
and
are denoted as the 'Entire' core in H2018-T 
and shown in their figure 1.
There was another observation (Obs 1)
 offset by about $3'$ and
with an exposure time of 49~ks.
Its total flux is fainter by a factor of about 3
and its spectral shape differs 
 in comparison to the other observations.
Therefore we do not use this offset data for this study.
Combined with the X-ray focusing mirror,
the SXS has a $3\arcmin\times3\arcmin$ field of view 
with an angular resolution of
$1\arcmin.2$ (half power diameter)
and covers the energy range of 2--12~keV.
We applied event screening based on the pulse rise time versus energy tuned for the wider energy coverage
and selected only the high primary grade events, in which the best spectroscopic performance is achieved. 
We applied pixel-by-pixel redshift correction and gain correction using a parabolic function,
as shown in appendix 1 of H2018-T.

\subsection{Basic spectral analysis methods}
The redistribution matrix file (RMF) 
and auxiliary response file (ARF) for spectral analysis are generated by \texttt{sxsmkrmf} and \texttt{aharfgen}, respectively.
We use two ARFs
for extended cluster emission and point source AGN components.
In H2018-T 
we examined three  separate ARFs
with different effective area calibration corrections, 
 finding
  the Crab ARF (tuned with the Crab spectra as defined in H2018-T)
  
as the best description of the data
for the \atomdb spectral model.
Accordingly, we use this ARF in the following analysis.

Energy bin sizes for
spectral and response (RMF and ARF) files
are all 1~eV.
No further binning is applied for the spectral fitting, 
except for plotting purposes.
Spectral fits are performed using the Xspec package (version 12.9.1h;  \cite{Arnaud1996})
with the modified C-statistic \citep{1979ApJ...228..939C}.
To model the cluster plasma emission, 
we use the atomic database \atomdb  version 3.0.9 \citep{smith01}.

\subsection{Backgrounds}
Non X-ray background (NXB; \cite{2018PASJ...70...18K})
spectra are produced from the night earth observations using \texttt{sxsnxbgen}
and used in the spectral fitting.
Table~\ref{tbl:nxb} lists identified and expected features in the NXB spectra.

The cosmic X-ray background is estimated to
 have an intensity of
$3\times 10^{-4}$ and
$3\times 10^{-5}$
in counts s$^{-1}$ keV$^{-1}$
over the SXS field of view
at energies of 2 and 10~keV, respectively.
These are well below 1\% and 0.15\% of the source counts at these energies.
We therefore ignore this background in the following analysis. 
The source and estimated NXB spectra are shown in figure 2 of H2018-T. 

\begin{table}[h]
  \tbl{Expected instrumental features in the SXS background spectra.}
{  \begin{tabular}{ccc}
  \hline 
Line \footnotemark[$*$] & Energy (eV) & note \\
\hline
Al Ka & 1486 & \\
Mn Ka & 5895 & Strongest. \\
Mn Kb & 6490 & \\
Ni Ka & 7470 & \\
Cu Ka & 8048 & \\
Au La  & 9671 & \\
Au Lb & 11510 & \\
Ag Ka & 22163 & Not clear in the NXB. \\
\hline 
Edge \footnotemark[$\sharp$]& Energy (eV) & note \\
\hline
Al K & 1559.6  & \\
Si K & 1839 & \\
Au M & \footnotemark[$\dag$] & Mirror, M1-5\\
Te (52) L & 4939 (L1) & Weak \\
Au (79) L & 11919(L3) & Mirror\\
Hg (80) M & 2295(M5), 2385(M4), 2847 (M3) & Escape edge energies \\
\hline
\end{tabular}
}\label{tbl:nxb}
\begin{tabnote}
\footnotemark[$*$] 
The line features are taken from \cite{2018PASJ...70...18K}. \\
\footnotemark[$\sharp$]
The edge energies
are taken from
the LBNL X-Ray Data Booklet\footnote{http://xdb.lbl.gov/}.
\footnotemark[$\dag$]
Au edge positions are
as follows, 2206(M5),
2291(M4),
2743(M3),
3148(M2),
3425(M1),
14353(L1), 13734(L2), 11919(L3)
in eV.\\
\end{tabnote}
\end{table}
\subsection{Analysis method - A blind line search}
\label{sect:fit-limit}

We search the SXS spectrum
for emission and absorption features
above the plasma model 
 over a broad range of possible
line position
following previous methods  used  with high energy resolution spectroscopy
(e.g., \cite{2007ApJ...656..129R}).
This is sometimes  referred to as a sliding box or blind search.
This method was also employed for the Perseus X-ray spectra
in \citet{tamura2015} and \citet{2017ApJ...837L..15A}.

\subsubsection{Initial test with a Line-free CIE model 
}

As an initial test to check  for prominent line features,
we use a single temperature CIE
(Collisional Ionization Equilibrium) 
model without any atomic line emission.\footnote{For this model we use SPEX with option "ion ignore". }
Spectral residuals are fitted with Gaussian line components 
 with a fixed intrinsic line width at 2~eV.
We detect  several strong (e.g., Fe He-line triplet)
and weak (e.g., He-like Cr) features.
All signals with statistical significance larger than 3$\sigma$
can be identified with atomic transitions expected from a 2--6~keV CIE plasma, 
as already shown in  published \hitomi papers.
No significant unidentified line emission is found.
In the Appendix, 
we show a spectral plot with annotated identifications (figure~\ref{plot-cp}).

\subsubsection{Baseline model - local-wide-band fits }
We determine baseline spectral models 
with physically reasonable parameters
to describe the data accurately
in the following wide and narrow bands fits.
Based on detailed spectral fitting in H2018-T 
we found the spectrum to be reproduced by 
two temperature CIE 
for the cluster  component and a power-law for the AGN emission, 
  along with AGN parameters obtained by \citet{H2018-AGN}.
These components are modified by a Galactic absorption column density
of $1.38\times 10^{21}$ cm$^{-2}$.

Instead of a multi-temperature model suitable for a wide energy band, 
we use single-temperature models ('tapec' in XSPEC)  
with changing parameters  tuned for each local energy band.
We fit the spectrum 
separately for energy ranges delimited at 
1.8, 2.7, 3.7, 6.4, 8.7, and 15.0~keV. 
The energy band from 6.25--6.32~keV
, which includes Fe-I emission from the AGN, 
is ignored in all cases.

The parameters determined at this step
are
temperatures (for continuum and for atomic lines),
metal abundances, 
cluster redshift,
velocity dispersion, 
and CIE normalization.
Other parameters are  held fixed.
In the 7.7--7.8~keV energy band
that includes both Fe and Ni strong line emission
we  allow Fe and Ni abundances to vary independently.
In other cases, a single metal abundance (represented by Fe) is  let free 
assuming solar abundance ratios.

\subsubsection{Gaussian Line flux - narrow band fits - }
We add a Gaussian line component ('zgauss' in XSPEC)
on top of the wide-band model
and determine the best-fit line flux allowing both positive (emission) and negative (absorption) values.

\begin{figure}[h]
  \begin{center}
\includegraphics[width=8cm]{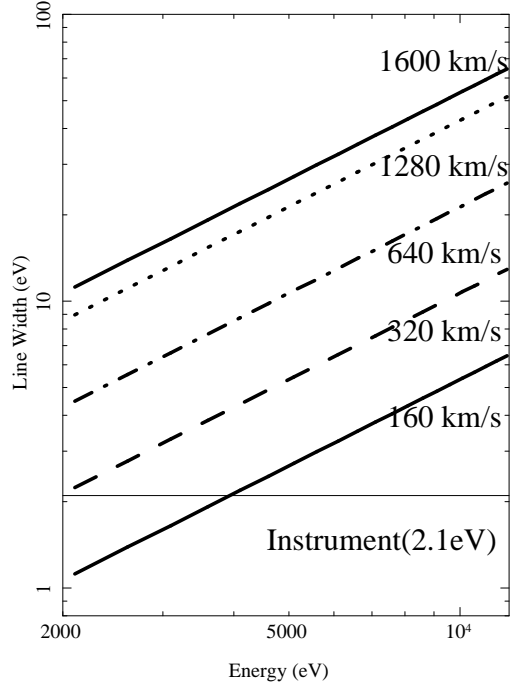}
  \end{center}
  \caption{
    The relation between the line energy ($E_{\rm Line}$) and
    intrinsic line width (\lgauss)
    for different velocity dispersion values ($V_{\rm dis}$), 
i.e., $\sigma_{\rm Line} = V_{\rm dis} / c \times  E_{\rm Line}$.
The horizontal line indicates the instrumental resolution at 2.1~eV.
  }\label{fig-e2s}
\end{figure}

Along with the line position $E_{\rm Line}$, 
the intrinsic line width (Gaussian sigma; \lgauss )
is an important but unknown parameter.
We first test the fitted parameters and method
with a fixed \lgauss $=3$~eV 
(subsection~\ref{ana-check} and \ref{res-flux:profile}).
Following this, 
we search for line signals from plasma or dark matter 
with energy-dependent \lgauss varying as $V_{\rm dis} / c \times  E_{\rm Line}$, 
where 
$V_{\rm dis}$ is a Doppler velocity dispersion of the emitter or absorber 
varying from 0 \kms to 1600 \kms with a step of 160 \kms, 
and $c$ is the speed of light.
Figure~\ref{fig-e2s} shows this relation.

At a plasma temperature of $\sim 4$~keV, 
atomic thermal broadening is 80 and 120 \kms for Fe and S ion lines, respectively.
For plasma lines,
 where will be an additional
turbulent velocity component of about 200 \kms, 
as measured by \citet{H2018-V}.
The line of sight velocity dispersion of the cluster galaxies 
is about 1300 \kms 
 \citep{1983AJ.....88..697K}.
The assumed $V_{\rm dis}$ brackets these velocity ranges.

Among the source (cluster and AGN) parameters,
only the cluster CIE component normalization is left free.
 Line signals are searched over a fixed energy grid  of 5~eV spacing. 
For each \Eline, 
a limited energy range ($E_{\rm Line} \pm $ \deltaEnar) is used. 
 
We use \deltaEnar as wide as 
$ 10 \times (\sigma_{\rm ins}^2 + \sigma_{\rm Line}^2)^{1/2}$, 
where $\sigma_{\rm ins}$ is the constant instrumental energy resolution of 2.1~eV.

For each \Eline, 
we determine
(1) a best-fit flux (the Gaussian component normalization),
and
(2) detection significance ($S$)
as defined as 
\begin{eqnarray}
S & = & \sqrt{\Delta C} = \sqrt{C_1 - C_0}, 
\end{eqnarray}
where $C_1$ and $C_0$
are C-statistic values
for no line and best-flux line models, respectively.
These values are assumed to
have a $\chi^2$ distribution.

\subsubsection{Fitting energy ranges and energy bin grids}
\label{ana-check}

We checked the effect of different energy ranges
(\deltaEnar) for the narrow band fit
on the line flux limit.
Changing \deltaEnar\,  from the default $\pm 50$~eV (for \lgauss = 3~eV)
to $\pm 25$~eV or $\pm 100$~eV 
had no significant effect on the resulting
distribution of $S$ 
or
line flux limits.

\begin{figure}[h]
\begin{center}
\includegraphics[height=0.4\textheight,angle=0]{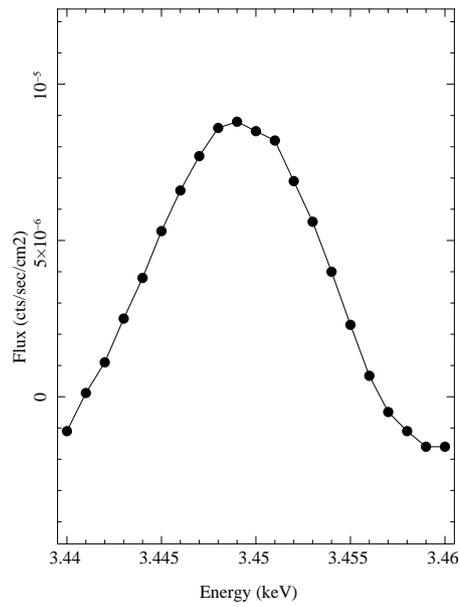}
\end{center}
\caption{
Best-fit line flux with a finer energy grid of 1~eV
at a limited energy range and \lgauss = 3~eV.
For our main results,
we use an energy grid of 5~eV spacing.
The plotted energy range represents a typical positive fluctuation.
The peak flux corresponds to
detection significance $S = 3.1$,
which is statistically insignificant considering a large number of trial energy bins
(see subsection~\ref{res-id:possible} regarding statistical evaluation).
This plot indicates that the effective width of line detection is wider than 2-3~eV
and that no new signal will be found by using an energy grid sharper than 5~eV.
See text for further explanation.
}\label{limit-plot2:1ev}
\end{figure}

We also checked the effect of  adopting a finer energy grid.
As an example,
figure~\ref{limit-plot2:1ev}
shows a result with an energy grid of 1~eV.
We find no significant variation over 5~eV energy bins, 
confirming that the 5~eV energy bin size is small enough
to detect a line feature with \lgauss $>3$~eV.

This limit is satisfied over most of the energy parameter space
and \lgauss $>320$\kms (see figure~\ref{fig-e2s};  $> 3$~keV).
Therefore 
energy grids finer than 5~eV
are expected to provide line search results consistent with the 5~eV ones.

\clearpage
\section{Results}
\label{sect:line-limit}

\subsection{Line flux limits}
\label{sect:line-flux-limits}
The wide-band fitting results and associated plasma emission models
are shown in figures~\ref{spec-e1:1}-\ref{spec-e4:last}.
The same data and similar fitting results
are given in
H2018-T and H2018-A.

\subsubsection{Residuals around atomic lines
}
The best-fit line temperatures for
the 1.8--2.7--3.7--6.4--8.7--15.0~keV bands
are
3.4, 3.6, 3.6, 3.9, and 3.9~keV, respectively.

  This increasing temperature trend
toward higher energy bands
reproduces the multi-temperature emission observed in H2018-T and other studies.
H2018-T found some differences
in line strengths 
between the best-fit single and two temperature CIE models,
particularly in
He-like S and Ar,
and Fe Ly$\alpha$ transitions.
These differences are up to 10--20\% in flux.
We find no significant residuals at these positions
(e.g., figures \ref{spec-e1:2}). 
Consequently, 
even if we  instead use two CIE models for the local bands,
we expect no significant improvement in goodness of fit,  
and hence no change in the line search results.

\subsubsection{Flux limit profiles}
\label{res-flux:profile}

\begin{figure*}[h]
\begin{center}
  \includegraphics[height=0.4\textwidth,angle=0]{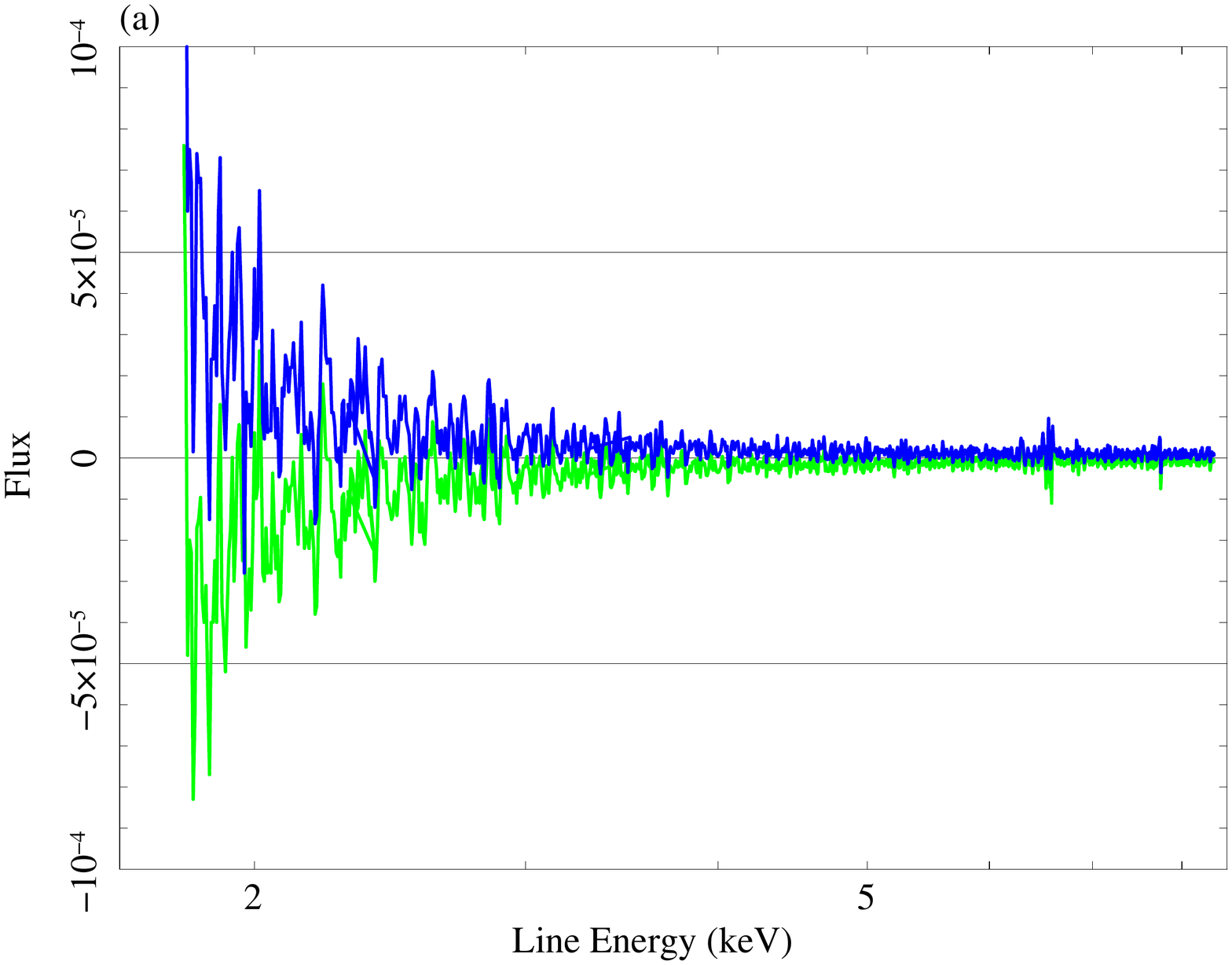}
\end{center}
\caption{
The line search result with \sigL = 3~eV.
Statistical 1$\sigma$ upper and lower limits
are shown
as different colored lines.
The y-axis units are in counts s$^{-1}$ cm$^{-2}$.
}
\label{limit-plot:flux1}
\begin{center}
\includegraphics[height=0.4\textwidth,angle=0]{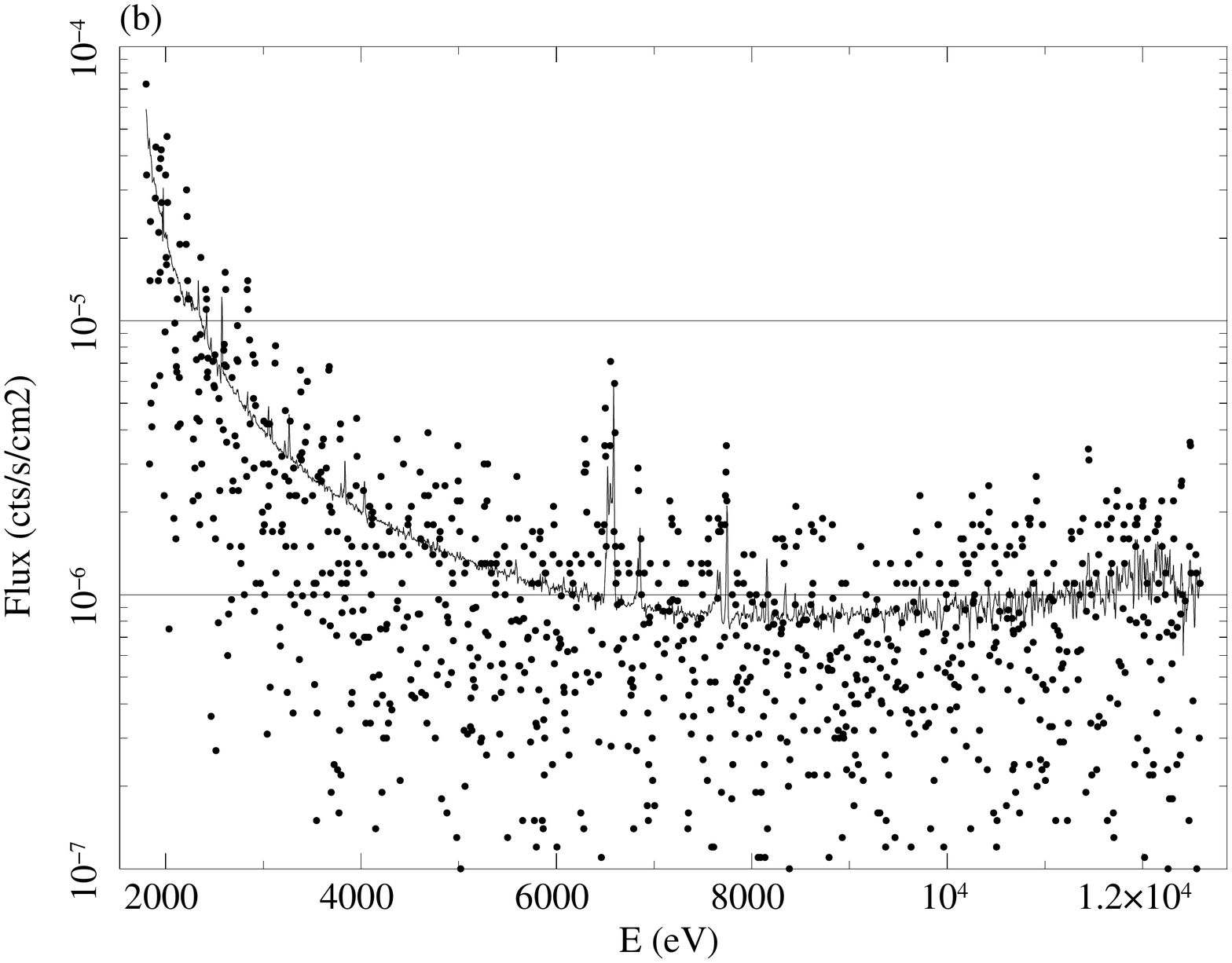}
\end{center}
\caption{
The best-fit flux (same data with figure~\ref{limit-plot:flux1}; $f_{\rm best}$)
and
positive 1$\sigma$ statistical error ($\Delta f_{+}$)
are shown by dot marks and the line,
respectively.
The latter is used as an estimate of the 1$\sigma$ upper limit flux of line {\rm emission}.
}\label{limit-plot:flux2}
\end{figure*}

A line search result with \lgauss = 3~eV  is shown in 
figure~\ref{limit-plot:flux1}.
The best-fit line fluxes ($f_{\rm best}$)
fluctuate statistically over a small range of energy bins between positive and negative values.
To estimate the 1$\sigma$ upper limit profile of line {\it emission} fluxes 
we use positive 1$\sigma$ statistical uncertainties ($\Delta f_{+}$).
Compared with these values,
standard upper limits
($f_{\rm limit} = f_{\rm best}+\Delta f_{+}$; including both emission and absorption)
are larger or smaller
by the $f_{\rm best}$ value which statistically fluctuate between positive and negative values.
Figure~\ref{limit-plot:flux2} shows $f_{\rm best}$ and the resulting upper limit profile of line emission.
This estimate is close to the standard {\it positive} upper limit averaged over energy ranges.

There are a number of line-like excesses in the upper limit flux profiles (figure~\ref{limit-plot:flux2}).
These are associated with strong plasma atomic lines in the source emission.
Strong source line emission increases background counts for additional faint line detection
and decreases sensitivity for the line search.

\begin{figure*}[h]
  \begin{center}
\includegraphics[width=0.5\textwidth]{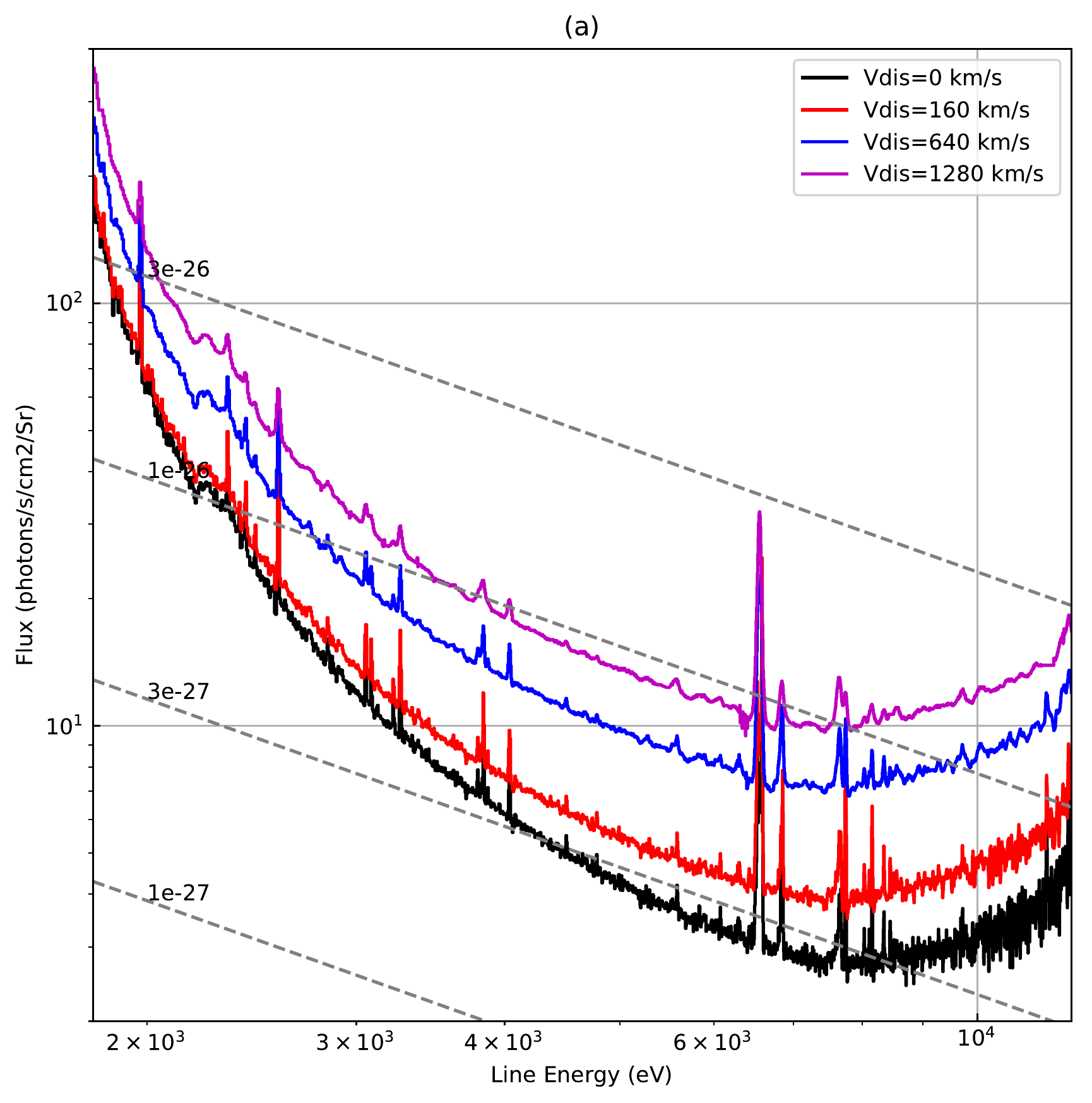}
\includegraphics[width=0.5\textwidth]{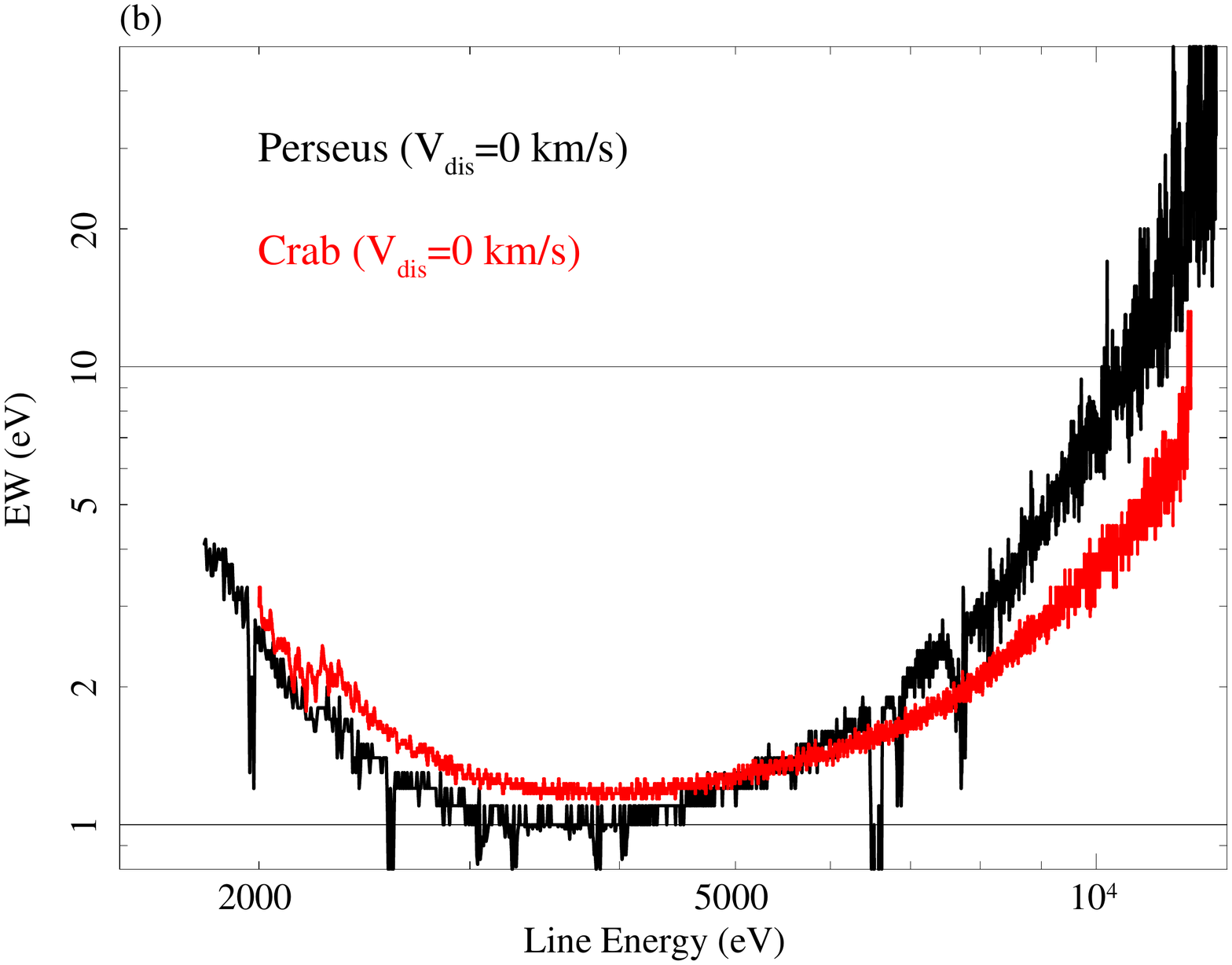}
\end{center}
  \caption{
    (a)
The estimated  3$\sigma$ 
upper limit flux
in units of
photons  s$^{-1}$ cm $^{-2}$ sr$^{-1}$ (Line Unit; LU).
Results with different \lgauss values are shown.
Limits with 
a number of different velocity dispersion values
are marked by different colored lines for a given sets of line energies.
Dashed lines indicate corresponding dark matter decay rate
in units of $s^{-1}$ given by equation~\ref{s-res:eq1}.
(b)
The line flux limits are converted
into line $EW$ relative to the source flux (cluster and AGN emission).
Results from the Crab observations \citep{2018-crab}
are also shown.
}\label{limit-plot3:lu}
\end{figure*}

Figure~\ref{limit-plot3:lu} 
 compares the results when
fixing $V_{\rm dis}$ at specific values.

Panel (b)
shows upper limits in unit of line equivalent width ($EW$)
 computed relative to
the source flux (cluster and AGN emission).
Absorption like features, e.g., at around 6.7~keV, 
are due to the source line emission.
These line flux limits
($f$)
can be converted into a decay rate
($\Gamma$)
for an arbitrary dark matter particle of mass $m_{\rm DM}$
by
\begin{eqnarray}
    f & \simeq & 9.3 \;
{\rm photons}\; {\rm cm}^{-2} {\rm s}^{-1} {\rm sr}^{-1}
  \frac{1}{(1+z)^3}
  \left(\frac{\Sigma_{\rm DM}}{10^3 M_\odot {\rm pc}^{-2}}\right)
  \left(\frac{\Gamma}{10^{-27} \; {\rm s}^{-1}}\right)
    \left(\frac{m_{\rm DM}}{{\rm keV}}\right)^{-1}, \label{s-res:eq1}
\end{eqnarray}
where $z$ and $\Sigma_{\rm DM}$
are the source redshift and
dark matter surface density.
To compute $\Gamma$ from $f$
we adopt
$\Sigma_{\rm DM} = 1750$ $M_\odot$ pc$^{-2}$
for the Perseus core
 \citep{2017ApJ...837L..15A}
and assume that emitted photons have an energy $m_{\rm DM} c^2 /2 $ as expected for sterile neutrino dark matter.

\subsubsection{Different line widths}
We examine flux limit profiles
 among models with different values of \lgauss (figure~\ref{limit-plot3:lu}).
Flux limits for the broader lines
are larger than those of the narrow ones 
in most energy ranges.
 These relative differences 
can be approximately
explained by
a difference in effective energy resolution combining instrumental and source widths.
When the line detection sensitivity
is limited by  background fluctuations,
its limit flux is approximately proportional to
the square root of the energy resolution.
In the actual limits based on spectral fitting,
the size of the fitting energy range,
\deltaEnar,
also affects the uncertainty of the continuum flux
and hence the line flux sensitivity.

\subsection{The Crab Nebula spectrum and systematic uncertainty
}
\begin{figure*}[h]
\begin{center}
  \includegraphics[height=0.90\textheight]{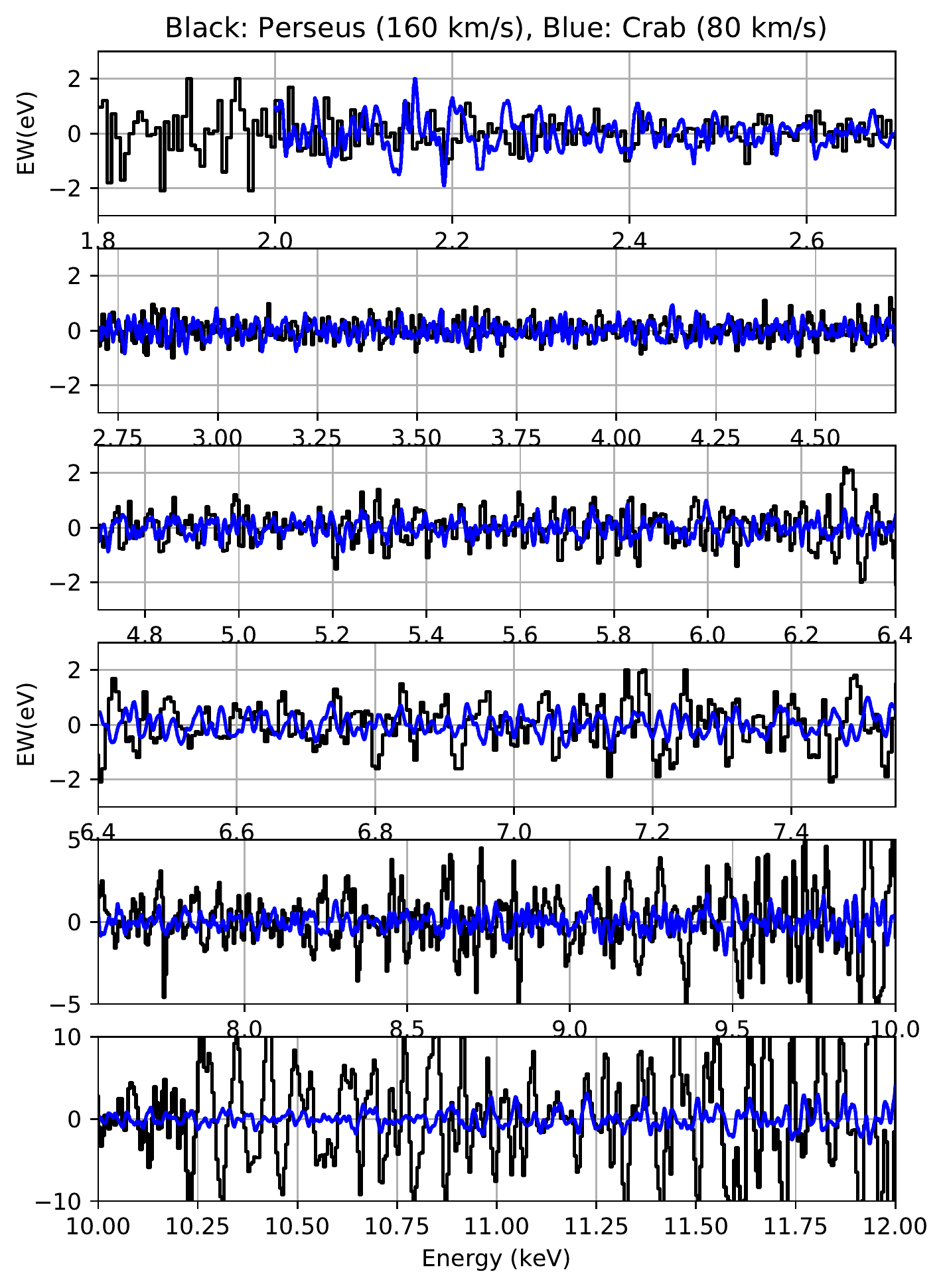}
\end{center}
\caption{
  Line search results 
in terms of the best-fit line intensity in EW (eV).
The Perseus and Crab results
are shown in black and blue colors.
For the Perseus and Crab fitting, 
respectively, 
\lgauss $=160$ \kms and 80 \kms.
}\label{plot-ew:1}
\end{figure*}

To evaluate systematic errors
from the effective area calibration,
we use the SXS Crab Nebula observations.
Using those observations, 
\citet{2018-crab}
performed a spectral line search similar to ours.
They used intrinsic line widths (\lgauss) fixed at a range of velocities 
from \lgauss $=$ 0 to 1280 \kms.
As shown in table~\ref{comp:cts},
the Crab photon counts are
about twice those in Perseus over the energy band.
The spectrum is intrinsically featureless
and hence cleaner than
 that of Perseus over most of the energy band.
These make the Crab data the best and only observations for this purpose.
Figure~\ref{plot-ew:1}
shows
comparisons of residuals in terms of best-fit line $EW$ 
from the Perseus and Crab spectra.

\begin{table}
\tbl{SXS observations of the Perseus core and Crab\footnotemark[$*$].}{
\begin{tabular}{cccc}
  \hline
  Energy (keV) & 2 & 4 & 10 \\
  \hline
  Perseus (cts/s/keV) & 0.05 & 0.4 & 0.02 \\
  (cts/eV/297~ks) & 15 & 120 & 4 \\
  \hline
  Crab (cts/s/keV) & 3 & 20 & 1 \\
  (cts/eV/9.7~ks )  & 30 & 200 & 10 \\
  \hline
\end{tabular}
}\label{comp:cts}
\begin{tabnote}
  \footnotemark[$*$]
  Counts of continuum emission and exposure times are given.\\
\end{tabnote}
\end{table}

The line flux  amplitude limits in the Crab spectrum
are smaller than 
 or comparable to  
those in Perseus over all energy ranges.
This is expected due to the better photon statistics in the Crab.
This allows us to use the Crab residuals
to estimate the systematic error associated with calibrations of the effective area and other instrumental features.

 In the Crab spectra,  
as presented in detail by \citet{2018-crab}, 
there are no residuals stronger than the typical statistical noise level
($<1-2$~eV in line $EW$).
We also find no clear positive or negative features common to both spectra.
These indicate that
any uncalibrated instrumental features
are too small to affect the line search significantly
within the statistical limit of the Crab observations.

From this comparison, 
we identify some line features
{\it brighter} than the Crab residuals
intrinsic in the Perseus spectrum.
One clear example is at around 6300~eV, 
which is the Fe-I K line from the AGN.
These can have astronomical origins
and are examined in the next subsection.

We check possible systematic errors associated with time variable response calibrations
by using only the Perseus Obs 3
 with an exposure time of about a half (146~ks) of the combined one (Obs 2,3, and 4).
With representative \lgauss values of 160, 640, and 1280 \kms, 
we find no new detection (detection significance $S>3.5$) nor 
significant change of $S$ over all energy bins.
In other words,
time variable calibration uncertainties
are too small to change the result from the combined observations.

\subsection{Line feature identification
}
\label{res-sig}

\subsubsection{Identified residuals}
\label{s-res:res}

As shown in the detection significance profiles
(figures ~\ref{spec-e1:1}-\ref{spec-e4:last}),
there are
 ranges of energy bins showing large detection significance (e.g., $S>3$) of emission or absorption.
Here we identify two systematic residuals
clearly associated with strong atomic line features.

(1) Around Fe \emissiontype{I} K at 6280-6300~eV:
These originate from the AGN 
as studied by \citet{H2018-AGN}.
 These are  caused by inaccurate modelling of the Fe\emissiontype{I} K line in the narrow band fitting.

(2) Around the strongest Fe He$\alpha$ complex at 6450-6620~eV: 
Those at lower energies (around 6500~eV)
may be due to inaccurate plasma thermal and/or atomic modeling
as examined in \citet{H2018-A}.
Those on the upper tail of the Fe He$\alpha$ resonance 
(around 6620~eV)
may be due to inaccurate modelling of plasma (Doppler) velocity structure
as discussed in \citet{H2018-V}.
Around this and other strong line features
possible calibration errors
of the detector response
can additionally contribute to observed residuals.

These systematic effects
make the detection of weaker features highly model-dependent  around strong atomic line features.

When the broader line widths ($>640$ \kms) are assumed,
these contaminations affect results around the wider energy bands.
We do not use the energy ranges given above for the line search.

\subsubsection{Possible line emission features}
\label{res-id:possible}

  Apart from the known systematic residuals stated above,
  there are energy bins showing large detection significance value
  ($S>3.2$)
as listed in Table~\ref{line-tbl:1}.
To identify line emission signals from the cluster plasma, 
lower significance ($S>2.5$) emission signals with \lgauss = 160 \kms or 320 \kms 
are additionally given.
Note that strong line velocities from the cluster plasma
were measured at 100--200 \kms using the same SXS data (e.g. \cite{H2018-V}).
In the appendix (figures~\ref{sf-vv:1}-\ref{sf-v160:320b}), 
zoom-in views of spectra around some of these marginal signals are shown.
Among them, there are two false signals at
1970~eV (Si) and 7755~eV (Fe),
which are
on modeled plasma transitions
and hence caused by  uncertainties on  those model emissivities.

The maximum values of $S$ associated with emission and absorption signals
are 3.5 and 3.6, respectively.
To evaluate these statistical significance considering the look elsewhere effect
(e.g., \cite{sekiya2015}), 
we assume a trial factor
 computed as the numbers of energy bins
[ $\sim$ (1.7--11.5)~keV divided by 5~eV]
times velocity bins (10),
which is about $2\times 10^4$.
By multiplying the Gaussian probability values by this factor, 
we predict
the number of false positive chances with $S>3.5$
to be 10 for each emission and absorption signal
over the searched energy and velocity space.
With this large trial factor,
a significant detection
corresponding to $>99$\% confidence (false positive of 0.01 )
requires $S > 5$.
Therefore any detected signals 
in the full energy bands
including those in Table~\ref{line-tbl:1}
 (not already associated with known systematics in subsection~\ref{s-res:res})
are consistent with statistical fluctuations.
We also confirm that
the $S$ distributions obtained 
are approximately consistent with those from statistical variations.
Based on these we conclude that we find no significant detection of any new line emission or absorption feature.

\begin{figure*}
\begin{center}
  \includegraphics[width=0.95\textwidth]{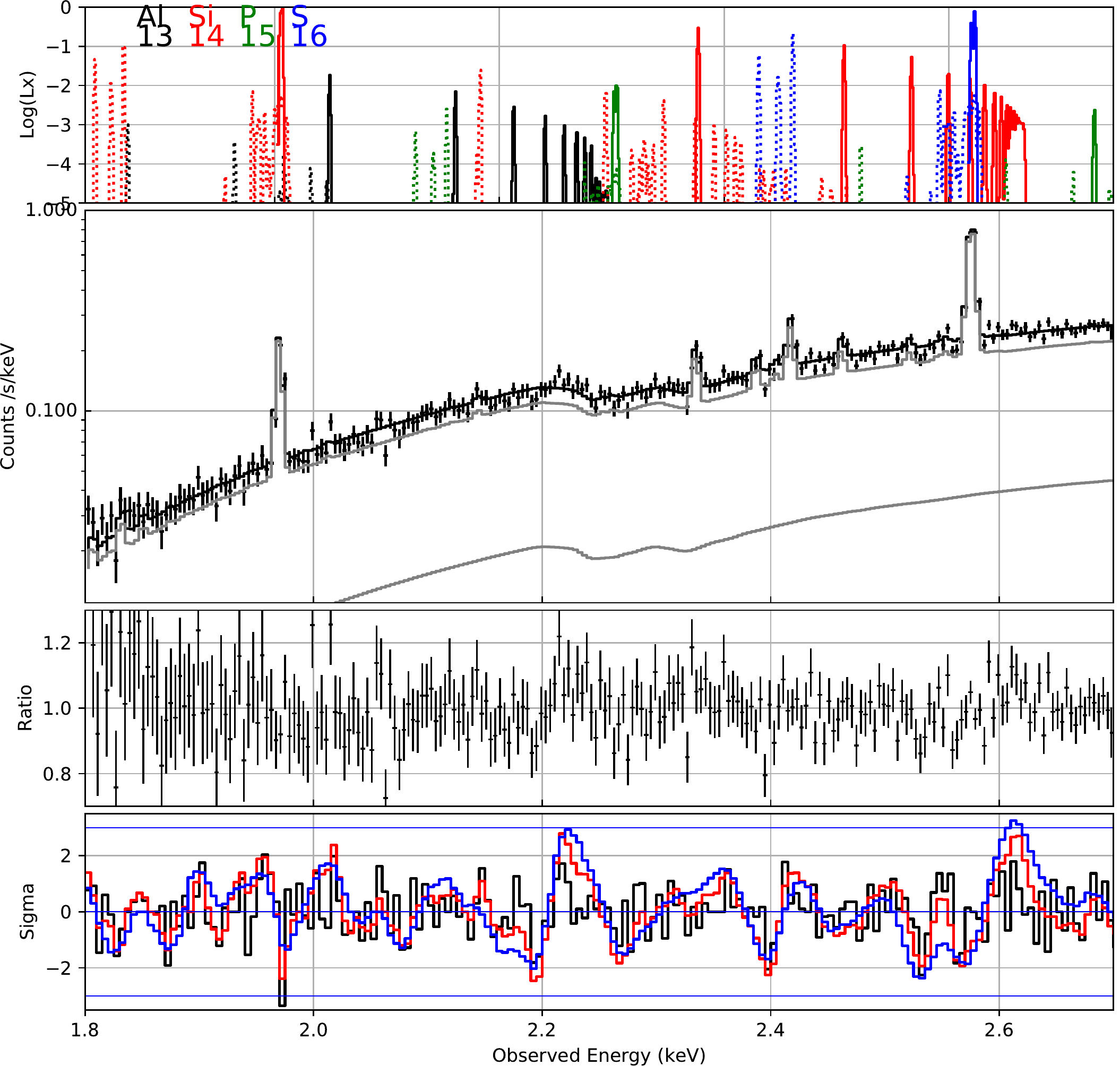}  
\end{center}
\caption{
(Top)
Line emission emissivity model in log scale
from a limited number of element ions are shown by different colors
as denoted by element names and numbers.
Different ionization emission are shown by different line styles,  
e.g., H-like (solid), He-like (dotted).
We used the SPEX model with temperature of 4~keV, 
the solar metal abudance ratio.
The x-axis (rest-frame energy) is shifted to match other panel ones (observed energy).
The line model in this panel
is similar to but not identical to 
the model used for the other panels.
 The energy bin size is 1~eV.
(Second panel)
The observed spectrum along with the best-fit local model (one-temperature APEC).
 For this and ratio plot, 
the energy bin sizes are 1--4~eV.
(Third panel)
The data to model (shown in the second panal) ratio.
(Bottom panel)
Line detection significance for each energy bin
in terms of 
($S=\Delta C)^{1/2}$ (in unit of $\sigma$).
Positive and negative values are for emission and absorption.
 Results with \lgauss = 160, 640, and 1280 in \kms 
are shown by
black,
red,
and blue
colors, respectively.
 The X axis is the energy grid of 5~eV spacing.
}
\label{spec-e1:1}
\end{figure*}

We searched for signals
by fixing \Eline and \lgauss
in grids of 5~eV and 160 \kms,
respectively.
Based on spectral fitting residuals 
and
$S$ profiles as functions of \Eline and 
\lgauss (figures~\ref{spec-e1:1}-\ref{spec-e4:last}),
no new and significant signal (e.g., $S > 5$) is expected
even when taking \Eline and  \lgauss
as free parameters.
We do not attempt to search for signals with
\lgauss $>1600$ \kms
[$> (20-100)$~eV],
where the SXS is less sensitive than previous CCD observations
with much larger grasp.

The signals listed above are not statistically significant detections 
but hints of possible features.
We provide possible identifications with atomic or instrumental features, 
guided by atomic line lists in H2018-A 
and the \spex package (version 3.03; \cite{1996uxsa.conf..411K}).
In the Appendix (table~\ref{line-pos})
we give a list of transitions
associated with these identifications.

We note possible detections
of atomic line emission
from high-$n$ to the ground shell emission 
in
Si \emissiontype{XIV}, 
Fe \emissiontype{XXV}, 
and 
Ni \emissiontype{XXVIII}
(table~\ref{line-tbl:1}).
These can be caused by charge exchange in the Perseus core.
Fe \emissiontype{XXV} associated with charge exchange 
in the same data set 
was previously reported in H2018-A,  
where another hint of detection from S \emissiontype{XVI} at $\sim 3.4$~keV (1.6$\sigma$)
was shown
along with a spectral analysis and discussion of the emission mechanism.

\subsubsection{Line absorption features}
Excluding the energy ranges of the strong plasma lines, 
mentioned in subsection~\ref{s-res:res}
and Si Ly$\alpha$ at 1970~eV, 
there are 
several absorption features 
with $S>3.2$ listed in table~\ref{line-tbl:1}.

Some of these are associated with
strong plasma 
or background lines.
 As mentioned above, 
these are not statistically significant
and too faint to examine further in regards to their possible origin.

The best-fit line fluxes 
fluctuate symmetrically between positive (emission) and negative (absorption) values
(figure~\ref{limit-plot:flux1}).
Therefore, upper limits on absorption strength (in flux or equivalent width)
would have a profile similar to those of the emission features 
shown in figure~\ref{limit-plot3:lu}.

\subsubsection{Cross-identifications with previous X-ray line searches}

For cross-identifications with our new potential line lists
(table~\ref{line-tbl:1}), 
we check previous line search results.
No cross-identified signal is found.

Using deep \suzaku observations , 
\citet{tamura2015}
searched for line features from the Perseus cluster.
Within the 2--6~keV energy band,
they found $>3\sigma$ signals at
3350, 4100, and 5810~eV.
In our list there is no signal close to these energies.
\citet{2010ApJ...725L.131P} reported a possible unidentified line at 8.7~keV in \suzaku Galactic center spectra.
\citet{2014arXiv1412.1170K}
reported fainter line-like feature at 9.4~keV and 10.1~keV
in \suzaku Galactic bulge spectra.
\citet{sekiya2015}
analyzed a large collection of \suzaku blank-sky observations (excluding the Galactic center)
in the 0.5--7~keV energy band and found line-like features with 2--3$\sigma$ statistical significance
at 600, 900, 1275,
4925,
and 5475~eV.
No coincident signal close to these energies is found in our search.

\begin{table*}
  \tbl{
 Possible 
line signal measurements and identifications. }
  {
\begin{tabular}{lllllll}
\hline
(1) Energy & (2) Energy &
(3) Velocity &
(4) Flux & (5) EW & (6) $S$ & (7) comments \\
eV   &  eV (rest)  & \kms &
& eV & $\sigma$ & \\
\hline
\multicolumn{7}{c}{$>3.2\sigma $ emission } \\
  2610 &   2655 & 1120-1280 &     3.7 &    3.4 &    3.2 & Si XIV high-n (CX, Nr=60), Figure~\ref{sf-vv:1} \\
  2840 &   2889 & 480 &     2.0 &    2.1 &    3.4 & excess of S He $\beta$ [Ap], Figure~\ref{sf-vv:1} \\
  7550 &   7680 & 960--1440 &     0.8 &    9.7 &    3.3 & Figure~\ref{sf-vv:1}\\
 10215 &  10392 & 0 &     0.2 &   15.3 &    3.5 & Close to the next one
 \\ 
 10910 &  11099 & 0 &     0.3 &   16.6 &    3.3 & Figure~\ref{sf-vv:1} \\ 
\hline
\multicolumn{7}{c}{$>3.2\sigma $ absorption} \\
  1970 &   2004 & 160 &   -10.0 &   -2.1 &   -3.4 & uncertainties with Si Ly$\alpha$ emission \\
  7755 &   7889 & 0 &    -0.4 &   -2.3 &   -3.6 & uncertainties with Fe line, Figure~\ref{sf-v160:e}\\
  8710 &   8861 & 0 &    -0.2 &   -3.8 &   -3.5 & Figure~\ref{sf-vv:1} \\ 
  8840 &   8993 & 0 &    -0.2 &   -3.9 &   -3.4 & \\ 
 10230 &  10407 & 160  &    -0.4 &  -10.5 &   -3.3 & near background feature (Figure~\ref{spec-e4:last}), Figure~\ref{sf-v160:320b}
 \\ 
 10655 &  10839 & 0 &    -0.2 &   -6.7 &   -3.2 & Figure~\ref{sf-vv:1} \\ 
\hline
\multicolumn{7}{c}{$>2.5\sigma $ and \lgauss = 160, 320 \kms emission} \\
  2840 &   2889 & 320 &     1.3 &    1.4 &    2.6  & see above \\ 
  3675 &   3739 & 320 &     0.8 &    1.5 &    2.9  & Figure~\ref{sf-v160:320} \\ 
  5260 &   5351 & 320 &     0.4 &    1.9 &    2.5  & Figure~\ref{sf-v160:320} \\ 
  7555 &   7686 & 320 &     0.3 &    3.9 &    2.6  & see above \\ 
  7735 &   7869 & 160 &     0.4 &    2.5 &    3.0  & Figure~\ref{sf-v160:e}\\
  8620 &   8769 & 320 &     0.4 &    6.8 &    2.7  & Figure~\ref{sf-v160:320}, Fe XXV (CX) \\ 
 10255 &  10432 & 160 &     0.3 &   11.6 &    2.6  & see above, Figure~\ref{sf-v160:e}
 \\ 
 10425 &  10605 & 160 &     0.3 &   13.9 &    2.8  & Ni XXVIII (CX, Nr=36),Zn XXIX (Nr=10), Figure~\ref{sf-v160:e} \\
\hline
\end{tabular}}
\label{line-tbl:1}
\begin{tabnote}
(2) Rest-frame energy with the cluster redshift. \\
(3) Line width in velocity.
(4) Line equivalent width. \\
(5) In units of $10^{-5}$ \UNITPFLUX.
A minus value in flux and 
detection significance
indicates an absorption signal.
(6) Detection significance.
This is 
single-line signal detection significance level
without correction for the look-elsewhere effect.\\
(7) In some cases, possible identification with atomic transitions are given.
See the Appendix (Table~\ref{line-pos})
for a list of possible identification transitions.
'Nr' in parentheses is the SPEX line number given in Table~\ref{line-pos}. 
See also spectral plot for expected emissivities for some signals.
Ap: Energies and other parameters are given in H2018-A. 
CX denotes charge exchange emission.
\end{tabnote}
\end{table*}

\begin{figure*}[h]
\begin{center}
  \includegraphics[width=0.95\textwidth]{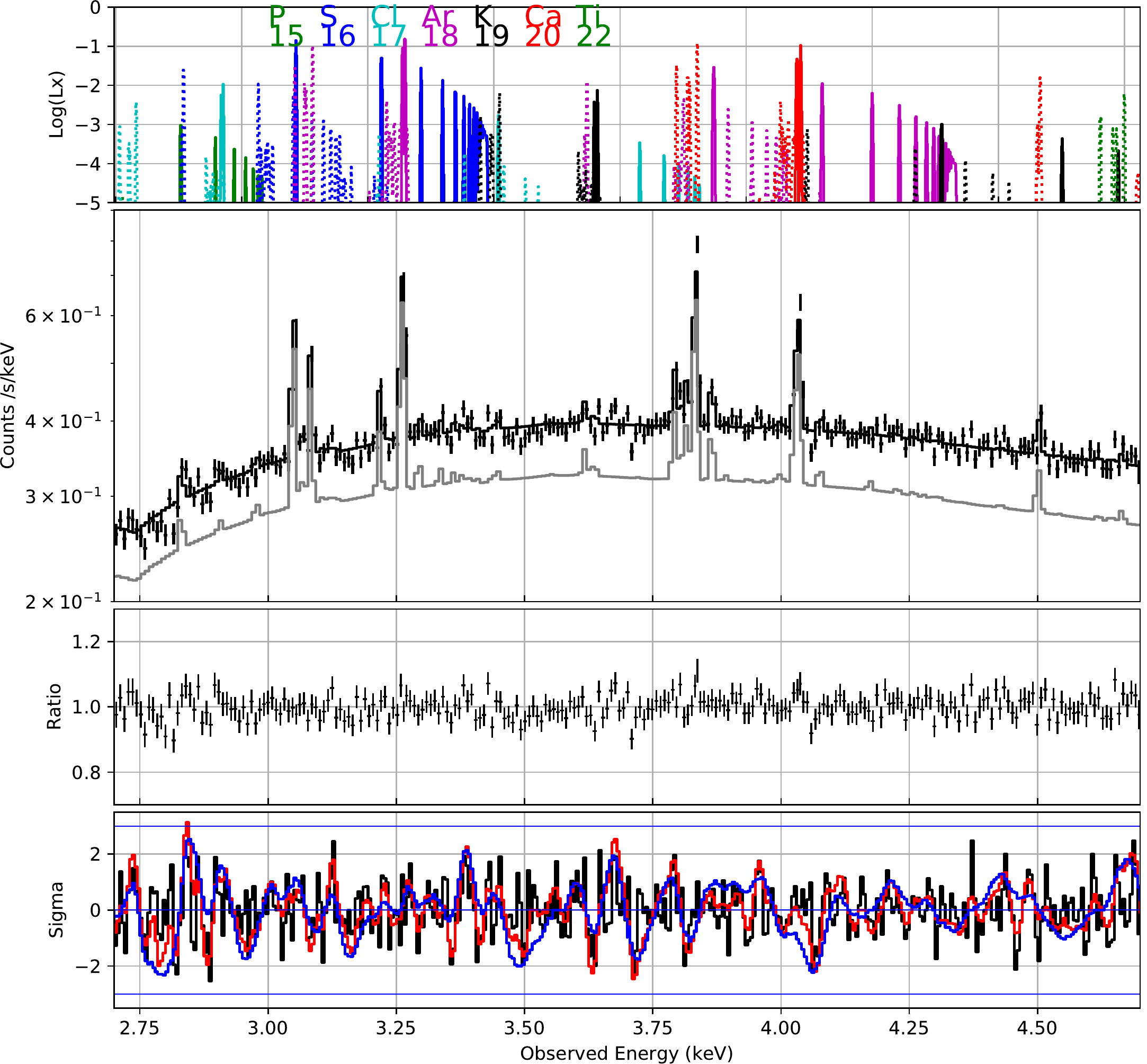}
\end{center}
\caption{Same plots as previous one,
  but for the energy range of 2.7--4.7~keV.
   The energy bin sizes for data and ratio plots are 1--8~eV.}
\label{spec-e1:2}
\end{figure*}

\begin{figure*}[h]
\begin{center}
  \includegraphics[width=0.95\textwidth]{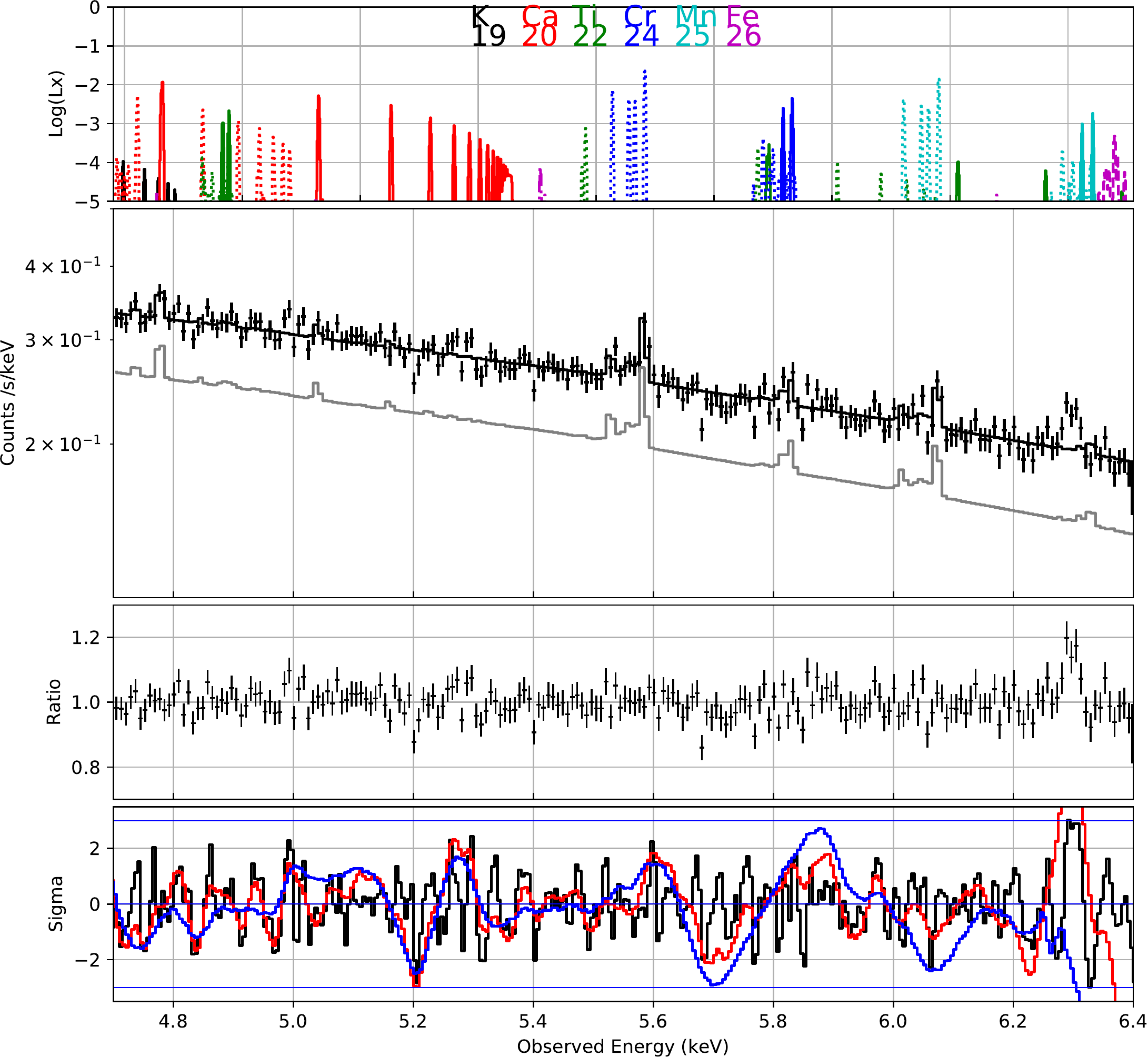}
\end{center}
\caption{Same plots as previous one,
  but for the energy range of 4.7--6.4~keV.
  The significant excess around 6.3~keV is originated from Fe \emissiontype{I} K emission of the Perseus AGN.
}\label{spec-e1:3}
\end{figure*}

\begin{figure*}[h]
\begin{center}
  \includegraphics[width=0.95\textwidth]{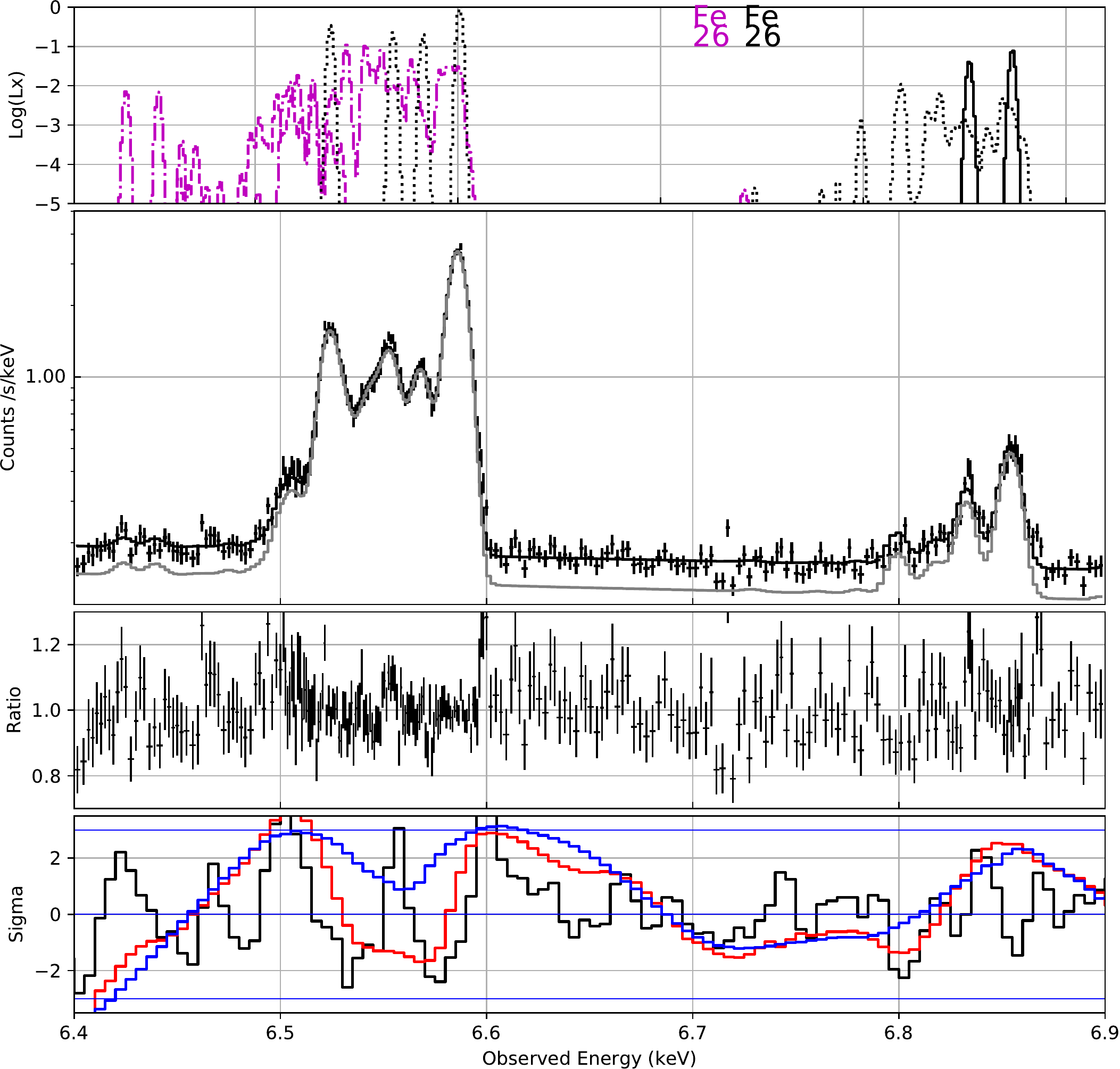}
\end{center}
\caption{
  Same plots as previous one,
but for the energy range of 6.4--6.9~keV.
}\label{spec-e1:4}
\end{figure*}

\begin{figure*}[h]
  \begin{center}
      \includegraphics[width=0.95\textwidth]{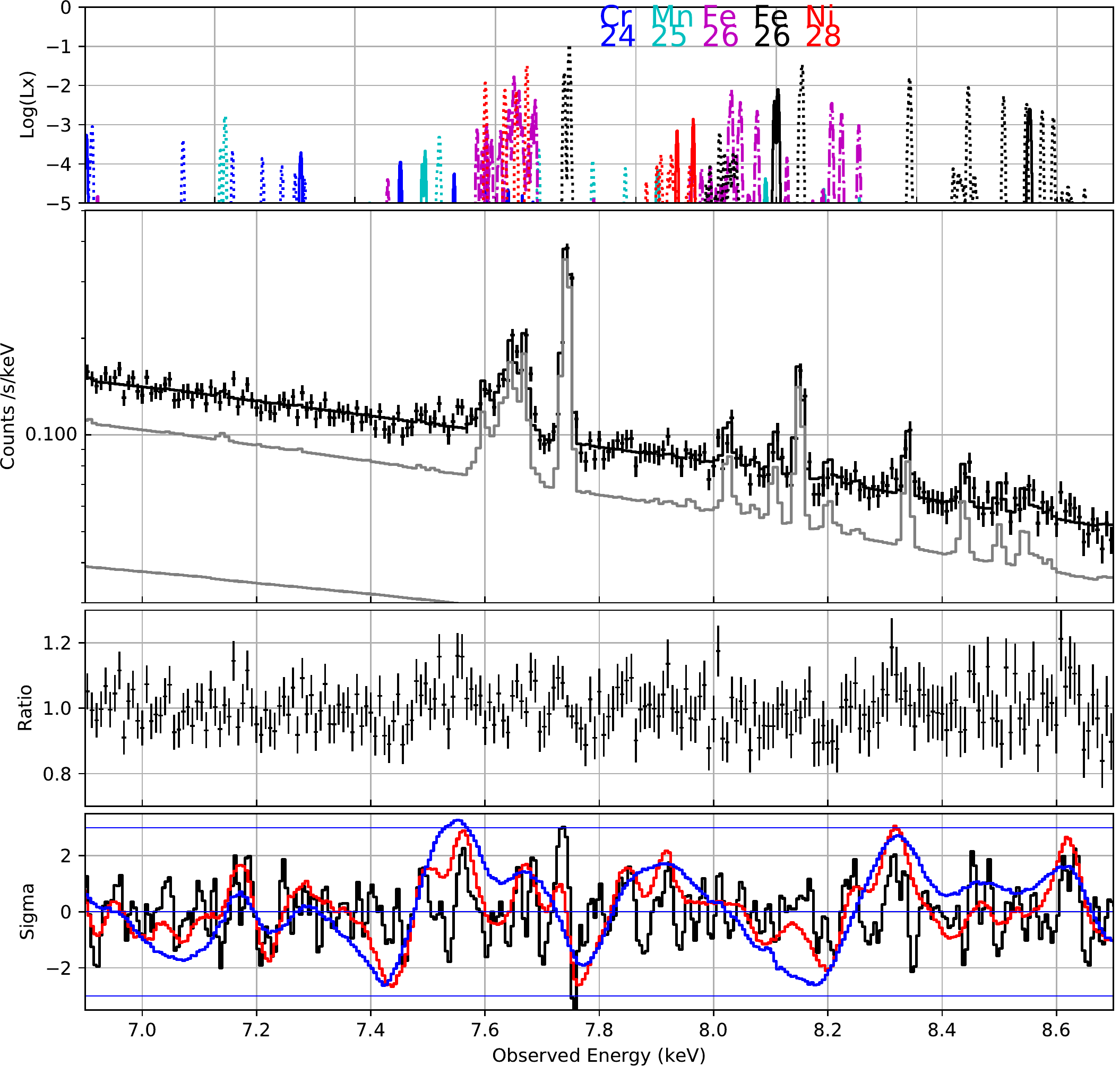}
\label{spec-e3:2}
\end{center}
\caption{
Same plots as previous one,
but for the energy range of 6.9--8.7~keV.
}\label{spec-e1:5}
\end{figure*}

\begin{figure*}[h]
\begin{center}
  \includegraphics[width=0.95\textwidth]{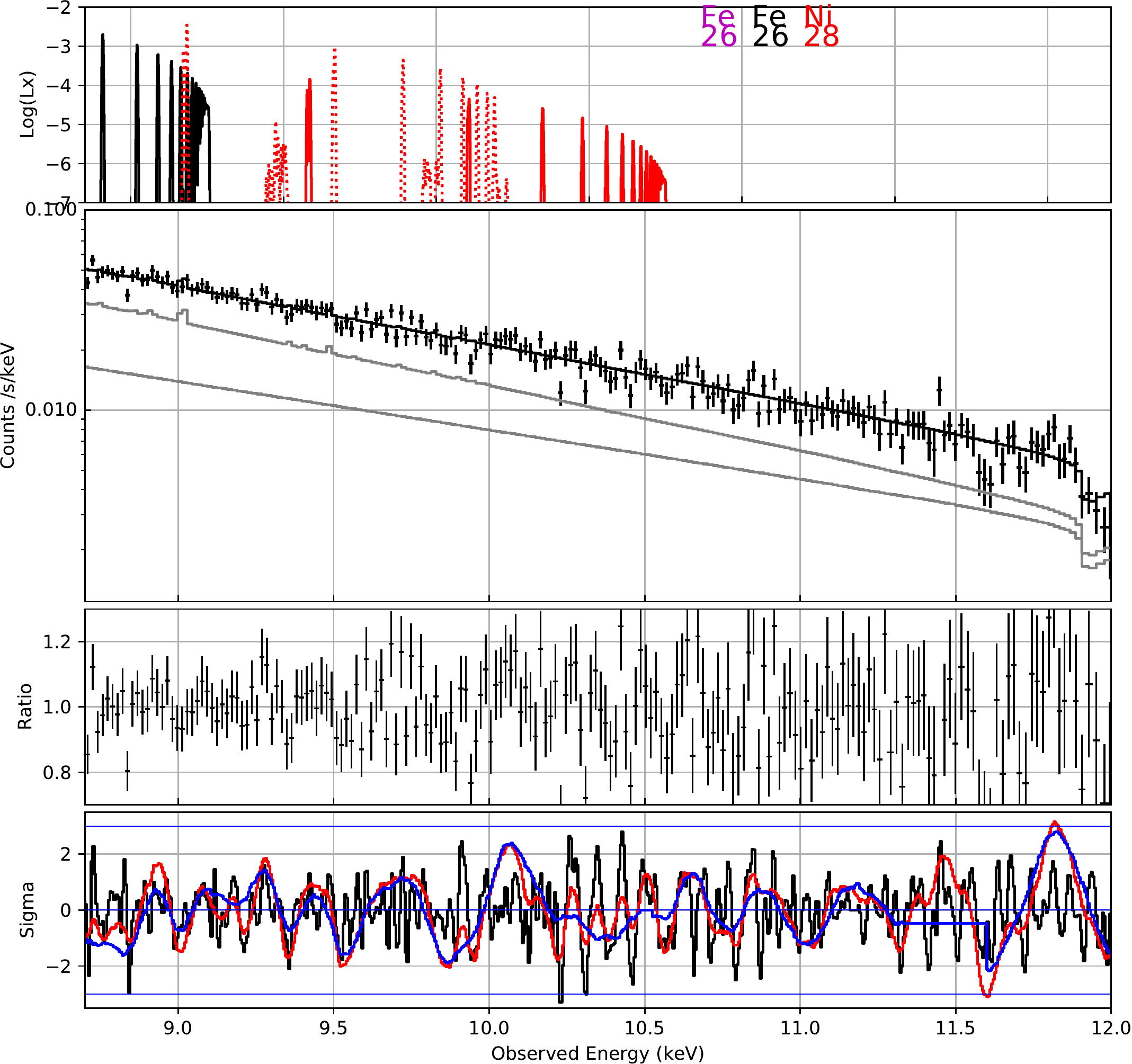}
\end{center}
\caption{
Same plots as previous one,
but for the energy range of 8.7--12.0~keV 
 and the energy bin sizes for data and ratio plots of 1--16~eV.
}
\label{spec-e4:last}
\end{figure*}

\clearpage
\section{Discussion}
\subsection{Comparisons with previous searches}
Here, we compare our line search results 
with previous studies.

\subsubsection{Absolute flux estimation}
To calculate absolute line flux and density values
(figure~\ref{limit-plot3:lu}),
we assume
(1) that all the observed photons originate from
the $3'\times 3'$ sky region corresponding to the SXS extraction area, and
(2) that the effective area may be approximated by the ARF simulated for a point source at the detector center.
The centrally-peaked surface brightness distribution
of the cluster 
and the angular resolution of $1'.2$ (half power diameter)
result in the photon spatial distribution
after the telescope mixing
to have an extension  not significantly wider than that of  point source.
The telescope vignetting reduction within the small field of view
is less than 5--10\% 
\citep{2018JATIS...4a1213I}. 
In addition,
the point source approximation for the ARF results in a flux to be $\sim 10$\% below that using the Chandra image by
\citet{2017ApJ...837L..15A}.
Considering these points,
we estimate our absolute flux uncertainty
to be less than 20--30\%, 
neglecting any additional error terms due to uncertainties 
e.g., in the pointing accuracy and stability and approximate treatment of the gate valve
 transmission \citep{Kelley2018}.

\subsubsection{Hitomi previous report around 3.5~keV \citep{2017ApJ...837L..15A}}
Using the same data set and similar analysis method,
\citet{2017ApJ...837L..15A} conducted a line search around 3.5~keV.
We compare the  published  result (their figure~3) with ours at the same energy, 
and find consistency in the upper limit profile.
\citet{2017ApJ...837L..15A}
identified 
an ``excess at 3.44~keV (rest-frame)''
at the position of a high-n transition of S\emissiontype{XVI}
and
the ``dip'' at 3.5~keV.
We confirm these features (figure~\ref{spec-e1:2}),
 but again at low significance.

\subsubsection{The SXS Crab search}

In figure~\ref{limit-plot3:lu}
we compare limits on the line flux equivalent width ($EW$) from the Crab \citep{2018-crab} 
\footnote{
For this plot,
line flux errors and continuum flux provided by M.Tsujimoto
are use to calculate the \EW limit.}
We calculate 1$\sigma$ uncertainties by taking the propagated error on the mean.
with the current Perseus result  by considering profiles of
statistical error in $EW$ ($\Delta {\rm EW}$).
For the background limited case
where number of line photon counts
($C_{\rm line}$)
are much less than the 
``background'' 
(i.e.,cluster continuum) counts ($C_{\rm bgd}$), 
i.e., $C_{\rm line} \ll C_{\rm bgd}$, 
\begin{eqnarray}
  \Delta {\rm EW} \sim \Delta C_{\rm Line}/C_{\rm cont}
  \propto \sqrt{2C_{\rm bgd}}/C_{\rm bgd} \propto C_{\rm bgd}^{-1/2}.
\end{eqnarray}
On the other hand,
in the photon limited case
where $C_{\rm line} \gg  C_{\rm bgd}$,
\begin{eqnarray}
  \Delta {\rm EW} \sim \Delta C_{\rm Line}/C_{\rm cont}
  \propto \sqrt{C_{\rm line}}/C_{\rm bgd} \propto C_{\rm bgd}^{-1}.
\end{eqnarray}
 The adopted spectral fitting method
may also affect the limits.
 
Below 6~keV, limits from the two observations are comparable.
Above this energy, 
the Perseus observations results in  larger 
\dEW
than Crab.

At higher energy,
the Perseus limit
increases
more rapidly than the Crab limit.
This is probably caused by
the photon limited factor,
$C_{\rm bgd}^{-1}$,
since for these energies
the Perseus counts are down to less than 10 counts per resolution element.

\subsubsection{The Suzaku Perseus limit}
\label{comp-xis}

\begin{figure*}[h]
\begin{center}
\includegraphics[height=0.6\textheight]{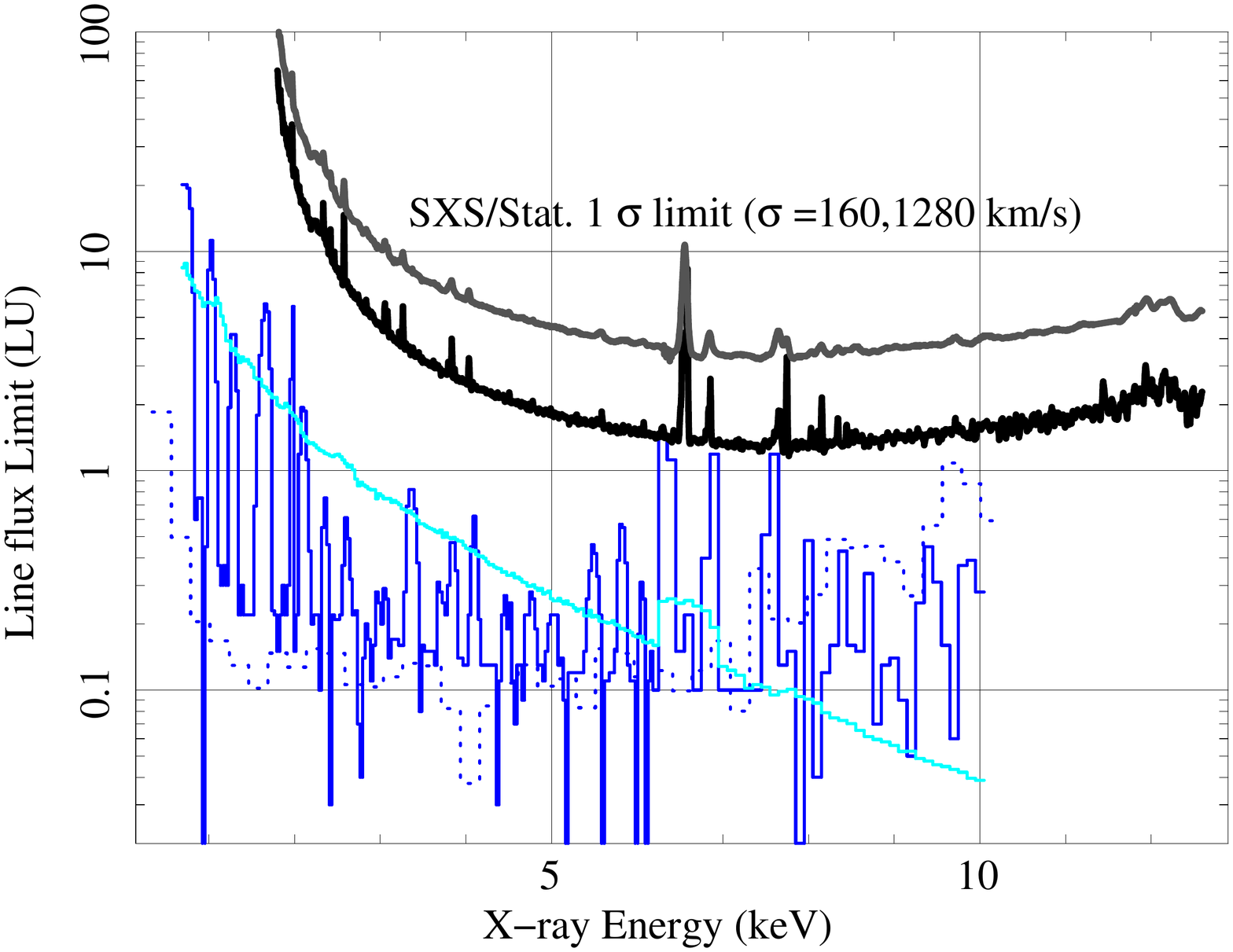}
\end{center}
\caption{
 
  Limits ($1\sigma$) on the line 
surface brightness 
in unit of photons cm$^{-2}$ s$^{-1}$ sr$^{-1}$ (LU).
The current SXS result ($1\sigma$) with \lgauss = 160 \kms and 1280 \kms 
(black and grey colors respectively; scaled from values in figure~\ref{limit-plot3:lu} and calculated in subsection~\ref{sect:line-flux-limits})
are compared with the \suzaku XIS results ($1\sigma$) from T2015.
The XIS limits refer to 
systematic  (1~eV in $EW$; light blue),
statistical (blue),
and for the Ursa Minor dwarf galaxy
(dotted line; \cite{Loewenstein2009}; ($3\sigma$)),
respectively.
}\label{comp-xis:p1}
\end{figure*}

We compare our SXS Perseus cluster results
with those from the \suzaku XIS (CCD) in \citet{tamura2015} (hereafter T2015).
Table~\ref{comp-tbl:tbl-exp}
compares the respective observational parameters.
Figure~\ref{comp-xis:p1}
shows line flux limits
from the two observations.

T2015 searched for line signals in the 2--6~keV energy range 
using spectra extracted from a $10'$ radius circle.
(see their figure~12 for the limit in $EW$).
They used the \suzaku Crab Nebula observation to correct their effective area
and estimated a systematic sensitivity to a weak line feature of less
than 2~eV in \EW from the maximum line flux in the Crab spectrum.

We note that in the XIS case 
the best-fit flux fluctuation amplitude 
is larger than
the statistical error
level over most energy bins.
Therefore,
the flux limit was determined by the systematic error estimated from the Crab spectra.
Contrary to the XIS case,
the SXS statistical errors are 
larger than the systematic one estimated from the SXS Crab observation
at least above
 6~keV
(figure~\ref{limit-plot3:lu}).
The Crab systematic errors 
are smaller ($<1$~eV) than those of the XIS.

In some energy ranges,
the current SXS observation
provides smaller $EW$ limits 
than XIS.
However,
this does not always translate into 
a smaller limit in line flux ($F_{\rm Line}$) or
intensity ($f_{\rm Line}$) as explained below. 
These values are related to the source flux intensity 
($f_{\rm source}$; the cluster and AGN flux over the relevant energy range in this case)
and observed sky area ($\Omega$) as follows.
\begin{eqnarray}
F_{\rm line} & = & EW \times F_{\rm source} .\\
f_{\rm line} & = & F_{\rm line} / \Omega 
             = EW \times F_{\rm source} / \Omega
             = EW \times f_{\rm source}.
\end{eqnarray}
Given the Perseus cluster surface brightness distribution (including the AGN flux)
and
$\Omega (SXS; 3'\times3') / \Omega (XIS; r<10') \sim 1/36$
(table~\ref{comp-tbl:tbl-exp}),
we estimate
$f_{\rm sou, SXS}/f_{\rm sou, XIS} \sim 12$ and 
$F_{\rm sou, SXS}/F_{\rm sou, XIS} \sim 3$.
Therefore
as given in the last equation above
for the same $EW$ limit,
the SXS $f_{\rm line}$ limit is larger than the XIS one by a factor of $>10$.
Even if the SXS \EW  limit is half of the XIS limit,
the SXS flux limit is 6 (12/2) times larger than the XIS one.

As shown in figure~\ref{comp-xis:p1}
the difference between the two observations
is consistent with the above estimation.
For the XIS spectra below 6~keV,
the effective area systematic error 
hinders the sensitivity to a weak line.
Above this energy range
the cluster Fe line emission,
instrumental features,
and continuum background
along with the decreasing effective area
reduce the sensitivity.
In contrast, 
the SXS high resolution spectroscopy
resolves and localizes these plasma and instrumental features,
resulting in less contamination around these features.
The decreasing SXS effective area dominates its sensitivity reduction.

\begin{table*}[bh]
  \tbl{\hitomi SXS and \suzaku XIS and observations.}{
\begin{tabular}{lllllllll}
\hline
Detector & Area $*$ & FOV $\dagger$ & exp$\sharp$ & Area$\times$ exp & Area$\times$ exp $\times$ FOV & E band \footnotemark[$\ddag$] & $\Delta E$ \footnotemark[$\S$]\\
         & (cm$^2$) & (arcmin$^2$) & (ks) & 
($10^6$ cm$^2 \cdot $ s) & 
$10^9$ cm$^2 \cdot $ s $\cdot$ arcmin$^2$) &
(keV) & (eV) \\
\hline
SXS & 100 & 9  & 289 & 29 & 0.26 & 2-12 & 5 \\
\hline
XIS/FI & 260 & 320 & 1040 & 270 & 86.5 & - & - \\
XIS/BI & 260 & 320 & 530  & 138 & 44.1 & - & - \\
total & 520    & -   & -    & 408 & 130.6 & 2--6 & 80-150 \\
\hline
\end{tabular}
}\label{comp-tbl:tbl-exp}
  \begin{tabnote}
    $*$ Effective area at energy of 3.5~keV.\\
$\dagger$ Detector's field of view.\\
$\sharp$ Exposure time. \\
\footnotemark[$\ddag$] Energy band  \\
\footnotemark[$\S$] Energy resolution in FWHM (full-width-half-maximum).
  \end{tabnote}
\end{table*}

\include{dis-dm}

\include{dis-cal}

\include{dis-tbl}
\include{conc}

\include{bib}
\include{app}
\end{document}

%% file: dis-dm.tex
\subsection{Limits on the dark matter decay rate}

Constraints on the X-ray flux of dark matter decay lines have been
derived using various targets and instruments with various band passes.
For example,
\citet{sekiya2015} used
a large quantity of X-ray diffuse background (XDB) data from \suzaku XIS
in a search for the dark matter signal from our Galaxy in the 1--7~keV energy range.
\citet{2017PhRvD..95l3002P}
used NuSTAR observations of the Galactic center in the 3--80~keV range.

To compare the sensitivity to a dark matter signal among different observations
, instead of the dark matter decay rate ($\Gamma_{\rm DM}$) itself,
we  use
the line flux density
($f_{\rm line}$)
divided by ($\Sigma_{\rm DM}$), 
which is
proportional to $\Gamma_{\rm DM}$, 
as given in equation~\ref{s-res:eq1}.
For the Perseus cluster 
$\Sigma_{\rm DM}$ within the SXS field of view
is calculated 
using the mass 
$(6-8) \times 10^{12}$ $M_\odot$ used in \citet{2017ApJ...837L..15A}.
Note that this and other 
$\Sigma_{\rm DM}$ estimations given below 
are uncertain up to a factor of two.

To estimate the line flux limit from SXS observations
we adopt the statistical errors,
which are larger than or comparable to the Crab systematic ones
(figure~\ref{limit-plot3:lu}).
Table ~\ref{dis-dm:tbl2} shows the resulting limits
along with previous analyses.
Compared with the XIS limit in T2015
for the same target but covering a larger area, 
current SXS limits
are larger (weaker)
at energy below 7--8~keV
and comparable 
at higher energies.

Because the cluster X-ray brightness is  more strongly concentrated toward the center
(roughly proportional to square of gas density)
than the estimated dark matter decay flux
(proportional to dark matter density),
observations limited to the central parts
such as the SXS one
have higher plasma flux (background)
and smaller signal-to-background ratio
than more extended ones such as the XIS observations.
This is a primary factor for the stronger limit on \dmDecay 
obtained by the XIS than the current SXS result 
at energies below 7--8~keV.
Above this energy range, 
the XIS sensitivity is limited by instrumental background
,
while SXS one is limited only by the decreasing effective area and photon statistics, 
as described in sub-subsection~\ref{comp-xis}.

For the energy range of 1--7~keV,
the 
XDB observations \citep{sekiya2015}
provides stronger limit on the line flux
and \dmDecay than other limits (table~\ref{dis-dm:tbl2}).
This is largely due to the much lower X-ray flux density in the XDB
than those in X-ray bright galaxies and clusters
as well as the larger solid angle and deeper observations.
Above about 7~keV,
this XIS observation suffers from the instrumental background
in the same way as the XIS Perseus observation in T2015.

For energies above 7~keV,
the NuSTAR observations of the Galactic center 
in \citet{2017PhRvD..95l3002P}
provides a strong limit on the line flux
and on \dmDecay.
This is largely due to a large solid angle of 
NuSTAR
(effective solid angle of $>10^4$ arcmin$^{2}$ 
and
detector area of 4-11 cm$^2$ as given in their table~1).
The large $\Sigma_{\rm DM}$
toward the Galactic center region
also contributes to the strong limit.

\begin{figure}[bh]
\begin{center}
 \includegraphics[width=0.8\linewidth]{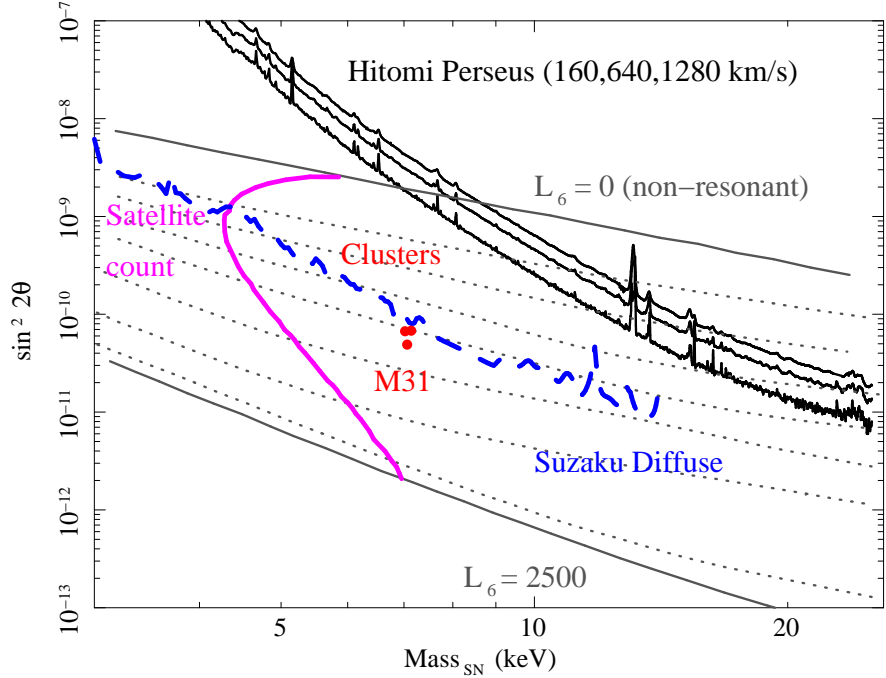}
\end{center}
\caption{
  Constraint from Hitomi observations of the Perseus cluster core
  are shown by the black solid line on the sterile neutrino mass-mixing angle plane.
The 3$\sigma$ line flux limit
with velocity dispersions at 160, 640, and 1280 in \kms
for lower to higher limits
and
the dark matter surface density,
$\Sigma_{\rm DM} = 1750$ $M_\odot$ pc$^{-2}$,
are used
along with equations in T2015
and a dark matter decay rate from
\citet{2010ApJ...714..652L}.
The $3\sigma$  limits
(correcting for the look elsewhere effect)
from the \suzaku X-ray diffuse background
spectra
are shown by the blue dashed line \citep{sekiya2015}.
The red data points indicate the claimed signals from
  M31~\citep{2014PhRvL.113y1301B} and stacked galaxy clusters~\citep{2014ApJ...789...13B}.
  Along the gray horizontal curves, the sterile neutrino relic abundance from the (non-)resonant production coincides with the observed dark matter mass density,
  for a given lepton asymmetry per entropy density,
  $L \times 10^{6} = L_{6} = 0, 4, 8, 12, 16, 25, 70, 700, 2500$,
  are shown.
  Here we utilize the public code \texttt{sterile-dm}~\citep{2016PhRvD..94d3515V} to calculate the relic abundance.
 The region of parameter space left of the magenta curve is disfavored by the satellite number count~\citep{2017PhRvD..95h3015C}.
}
\label{app-dm}
\end{figure}

Our flux limits along with the assumed 
$\Sigma_{\rm DM}$
are translated into those on the sterile neutrino mass-mixing angle plane
as shown in figure~\ref{app-dm}.

%% file: dis-cal.tex
\subsection{Instrumental calibration and atomic line emission models}
Our analysis confirms
importance of the instrumental calibration
to maximize its spectroscopic capability.
Firstly,
in our analysis, 
effective area calibration on a fine energy grid
limits the systematic uncertainty of weak line fluxes.
This is particularly crucial
with bright X-ray sources like galaxy clusters and the Galactic center.
Without the \hitomi observations of the Crab Nebula,
a line free bright X-ray source,
we could not examine the effective area uncertainty.
Secondly,
the energy scale and response
should also be stable and calibrated well for
separating source features
from statistical noise and atomic or instrumental features
without a-priori knowledge of  position and Doppler velocity.
In the \hitomi case, 
these were calibrated without
the originally planned per-pixel, simultaneous X-ray energy
calibration using the Modulated X-ray Source.
For the new mission, 
this system will be used to improve spectroscopic performance.
See e.g., \citet{2018JATIS...4a1214K}
for the SXS calibration.

In addition to the instrumental features, 
atomic line emission parameters 
should also be calibrated and improved. 
The obtained lack of additional lines
(subsection~\ref{res-sig})
supports
the accuracy level of the current plasma modeling, 
as examined in H2018-A. 
As shown in H2018-A
and this study (figures~\ref{spec-e1:1}-\ref{spec-e4:last})
a forest of weak atomic lines
are expected to appear depending mainly on the source plasma temperature and metal abundances.
Most weak features have not yet been observed and hence have uncertainties
in their positions and emissivities.
Uncertainties in the plasma emission modeling also
limit the identification of new features.
In fact the SXS Perseus spectra
suggest the presence of charge exchange emission on top of the thermal emission
(subsection~\ref{res-sig}).
Once these atomic features are
resolved and modeled accurately,
 the analysis should become more suitable for carrying out blind searches.
Alternatively, these atomic contamination  features
may be avoided by observing X-ray faint sources
that have lower dark matter densities but better dark matter to plasma emission background ratios.

%% file: dis-tbl.tex
\begin{table*}
  \tbl{Limits on the dark matter signal.
    }{
\begin{tabular}{llrlrrrr}
\hline
Origin & Target & 
$\Sigma_{\rm DM}$ \footnotemark[$*$]  & Ins. \footnotemark[$\dag$] & 
Energy & f limit \footnotemark[$\ddag$] & 
$f/\Sigma_{\rm DM}$ & $\Gamma_{\rm DM}$ \footnotemark[$\|$]\\
    &        & 
   &      & 
keV    & ($1\sigma$; LU) & 
& $10^{-28}$ s$^{-1}$\\
\hline
This (\lgauss= 640 \kms)   & Perseus & 1750   & SXS & 2                 & 30   &  170  &  80  \\
 --    & --      & --     & --  & 5                 & 3.0   &   15  &  20  \\
 --     & --      & --      & & 8                 & 2.0   &    12  &  20  \\
 T2015  & Perseus (R$<10'$) & 800   & XIS & 2     & 2.0   &   25  &  10  \\  
--     & --                 & --   & --  & 5     & 0.3   &    4  &   4  \\  
--     & --                 & --   & --  & 8     & 0.5   &    6  &  10  \\  
\citet{2014ApJ...789...13B} & Perseus & 820 & XMM/MOS & 3.5 & 1.9 &   23  &  18  \\ 
\citet{sekiya2015} & XDB & 30 & XIS & 1--7 & $3\times 10^{-3}$ &    1.0  &   1  \\  
\citet{2017PhRvD..95l3002P} & GC   & 300 & NuSTAR & 3-7 & $2\times 10^{-1}$ &     7  &   7  \\ 
 --     & --      & --      &           & 7-100 & $2\times 10^{-2}$ &    0.7  &   7  \\  
\hline 
\end{tabular}
      }
      \label{dis-dm:tbl2}
\begin{tabnote}
\footnotemark[$*$] Dark matter surface mass density in units of $M_\odot$ pc$^{-2}$.\\
\footnotemark[$\dag$] Mission and instrument name.\\
\footnotemark[$\ddag$]  X-ray line flux $1\sigma$ limit. We assume a Gaussian distribution for the error range. \\
\footnotemark[$\S$]  Given in units of $10^{-4}$ LU pc$^2/M_\odot$.  \\ 
\footnotemark[$\|$]  Dark matter decay rate limits corresponding to the line flux limit as in equation~\ref{s-res:eq1}. \\
\end{tabnote}
\end{table*}

%% file: conc.tex
\subsection{Conclusion and Future Prospects}
\citet{2017ApJ...837L..15A}
used the SXS high energy resolution spectrum
and
did not confirm the 3.5~keV signal reported in \citet{2014ApJ...789...13B}.
We extended this search into the full observed energy band of 2--12~keV
We have demonstrated the unique advantages of
high resolution spectroscopy for weak line searches
and provided a pilot study for future reference.

We provide lists of possible 
 line emission as well as line absorption features (table~\ref{line-tbl:1}).
 These signals may represent
statistical fluctuations or, in some cases, 
the first hints of unmodeled atomic lines or lines of other origin.
These should be examined with
future laboratory and cosmic  X-ray experiments.
Hint of faint line features possibly originating from
Zn
and charge exchange emission
demonstrates
the vast potential of high resolution spectroscopy of cosmic plasma.
These features 
will be resolved by deep and targeted observations
and be unique probes for cosmic plasma.

We found no significant unidentified line emission nor absorption
(section~\ref{res-sig}).
The line flux upper limits as functions of
energy and their dependence on intrinsic line width
are displayed in figure~\ref{limit-plot3:lu} and table~\ref{dis-dm:tbl2}.

Due to 
the smaller grasp
and shorter exposure 
of the current observation 
compared with those of previous searches,
our flux limit is not largely stronger than those claimed previously.
However,
we should examine carefully the previous analyses based on lower energy resolution spectra.
As discussed in section 1
some original claimed detected fluxes of a 3.5~keV line (\cite{2014ApJ...789...13B}) are in conflict with other analyses, 
including those from high energy resolution spectra.
This may indicate that some of these reports underestimated systematic uncertainties for weak lines.
Accordingly, our results are complementary to past studies for X-ray line searches, 
utilizing a target with one of the highest dark matter column densities
in the nearby universe.

The small grasp of the SXS is an impediment to measurement of fluxes of
possible unidentified faint signals.
Even ignoring background and other instrumental systematic errors,
to detect and resolve the signal
more robustly than previous claims, 
at least 10 line photons are required.
This approximately corresponds to the $3\sigma$ photon-limit sensitivity
as estimated in \citet{2014arXiv1412.1176K}
with the designed SXS capability.
The expected sensitivities
with 1~Ms exposure
are about
0.3 (0.5~keV),
0.06 (1--5~keV),
and
0.1 (10~keV),
all in LU at energies given in parentheses, 
for a narrow line emission.
These fluxes are not significantly lower than the level
constrained previously
(see table~\ref{dis-dm:tbl2} for examples).
Additionally lines  can possess a variety of intrinsic widths and 
  can be particularly broad 
in massive objects.
Therefore, 
to detect such faint unidentified signals, 
much deeper data 
or larger grasp with good energy resolution
is required.

The next opportunity
for astronomical high resolution non-dispersive X-ray spectroscopy 
will be the calorimeter on the JAXA-NASA Hitomi recovery mission,
\xarm \citep{Tashiro-xarm}.
This will have basic design and capabilities inherited from \hitomi, 
and will observe a variety of dark matter-rich systems.
With \xarm,  the energy range will be extended down to 0.5~keV
and much longer observation times employed.
\citet{2014arXiv1412.1176K}
discussed detailed strategies
for the dark matter search using SXS-type spectroscopy.
If any unidentified candidate signals
are suggested by theories or by X-ray and other observations,
the \xarm instrument will be the first 
to identify or reject those signals.

Acknowledgement.
We thank T.Takahashi and the \hitomi team members for their work.
This work follows pre-launch study of dark matter search and
early collaboration for the 3.5~keV signal by the \hitomi SWG members. 
Comments from
F. Paerels,
N. Werner,
J. Hughes,
S. Ueda,
and R. Blandford are helpful.
We thank the referee for useful suggestions and comments.
We acknowledges support from Grant-in-Aid for Scientific Research from the MEXT,
JP17K05393 (KS), 18K03704 (TK), 
the RIKEN Special Postdoctoral Researcher Program (SN),
Yamada Science Foundation (TT), 
ERC Advanced Grant Feedback 340442 (ACF, CP), 
IBS under the project code IBS-R018-D (AK), 
STFC ST/R000506/1 (PG)
and
the Programma per Giovani Ricercatori - anno 2014
``Rita  Levi Montalcini'' (FT).

%% file: app.tex
\appendix
\section{Associated tables and plots}

\begin{figure}[h]
\begin{center}
\includegraphics[height=0.40\textwidth]{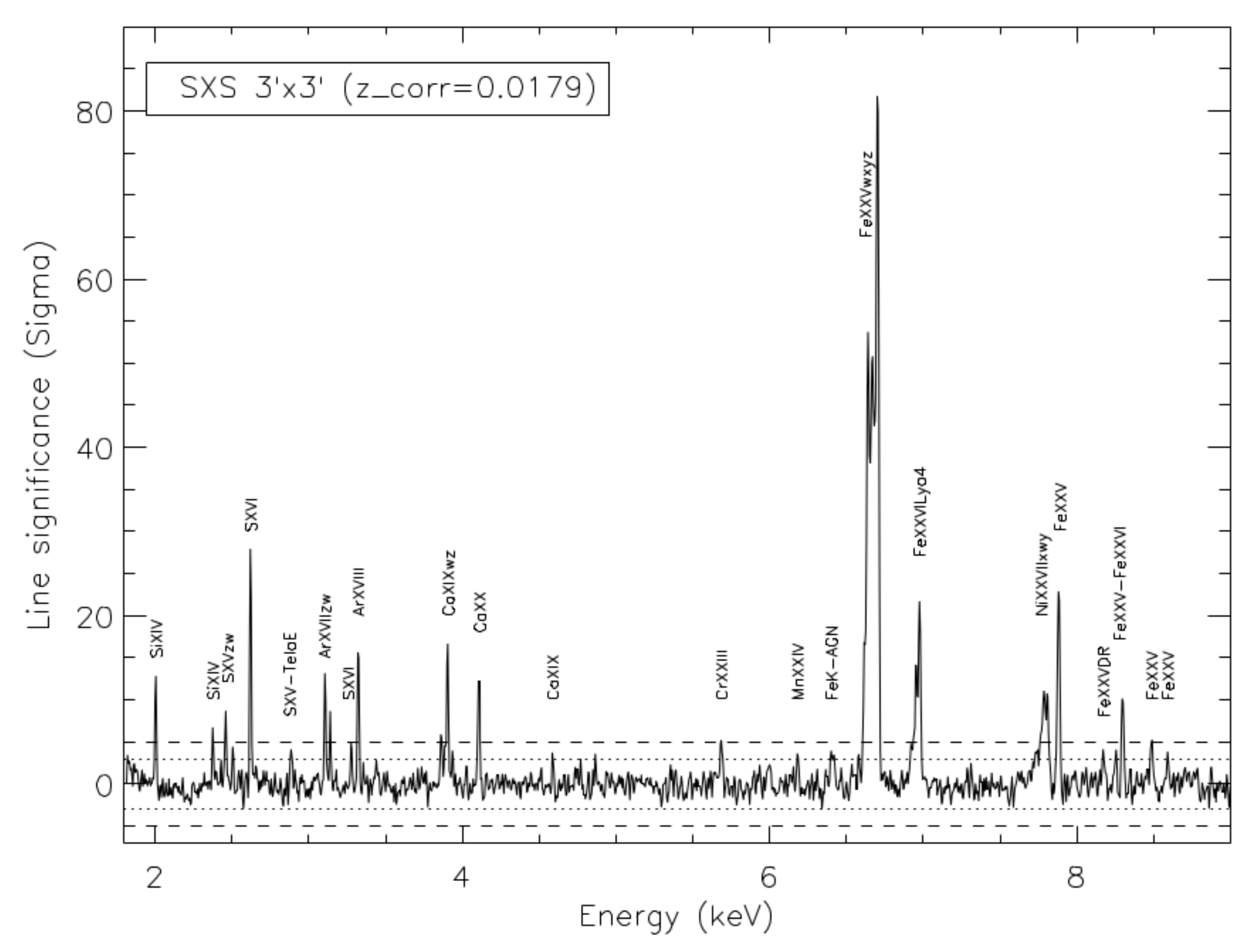}
\end{center}
\caption{
  The Perseus core spectrum
  and detected line emission signals.
  The vertical axis is given by line detection significance in $\sigma$.
  The continuum emission is modeled by a single temperature CIE model and subtracted.
  Atomic line identifications are shown.
  Dashed and dotted horizontal lines show 5 and 3 $\sigma$ detections.
}\label{plot-cp}
\end{figure}

\begin{table*}
\tbl{ Possible identification and associated ines from the SPEX database.}
{
\begin{tabular}{llllll}
\hline
(1) Nr & (2) ele & (3) stage & (4) Energy & (5) Emi & (6) Up \\
\hline
    48 & Si &    XIV &  2646.5 &  1.1e-03 &       10p \\ 
    54 & Si &    XIV &  2651.1 &  7.9e-04 &       11p \\ 
    57 & Si &    XIV &  2654.6 &  3.0e-04 &       12p \\ 
    60 & Si &    XIV &  2654.6 &  6.0e-04 &       12p \\ 
    66 & Si &    XIV &  2657.4 &  4.7e-04 &       13p \\ 
    72 & Si &    XIV &  2659.6 &  3.7e-04 &       14p \\ 
    78 & Si &    XIV &  2661.3 &  3.0e-04 &       15p \\ 
    30 & Ni & XXVIII &  10556.4 &  7.9e-06 &        7p \\ 
    33 & Ni & XXVIII &  10607.8 &  2.6e-06 &        8p \\ 
    36 & Ni & XXVIII &  10607.8 &  5.1e-06 &        8p \\ 
\hline
\end{tabular}
}\label{line-pos}
\begin{tabnote}
(1) Nr: Identification number defined in the SPEX database. \\
(2) Ele: Element name. \\
(4) Energy: rest frame energy in eV. \\
(5) Emi: emissivity for a 4.3~keV plasma with a certain emission measure normalizaed by
that of Fe XXV resonance (the strongest one). \\
(6) Up: Upper level electron configulation. \\
\end{tabnote}
\end{table*}

\begin{figure*}[h]
  \centerline{\hbox{
      \includegraphics[width=0.330\textwidth]{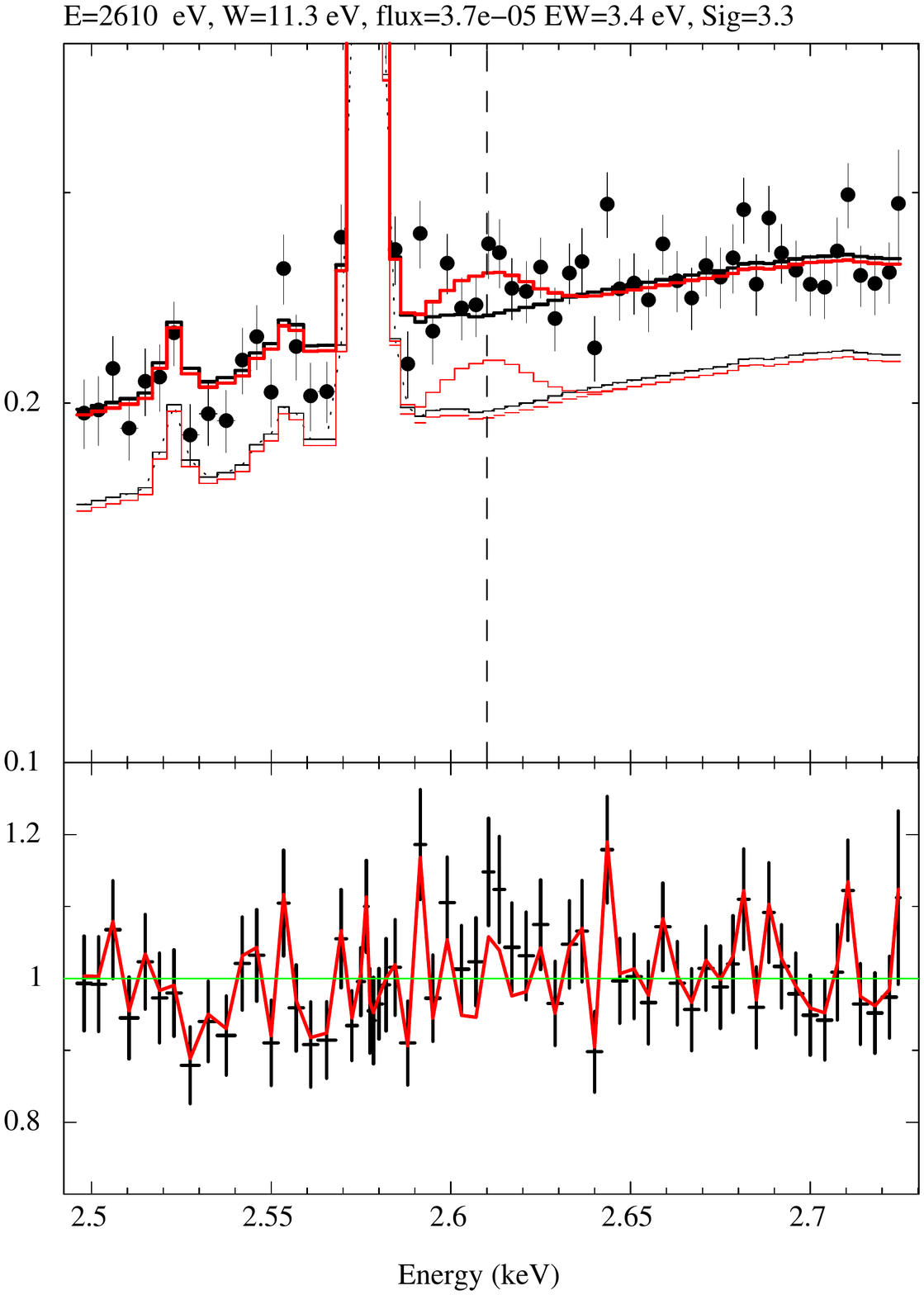}
      \includegraphics[width=0.330\textwidth]{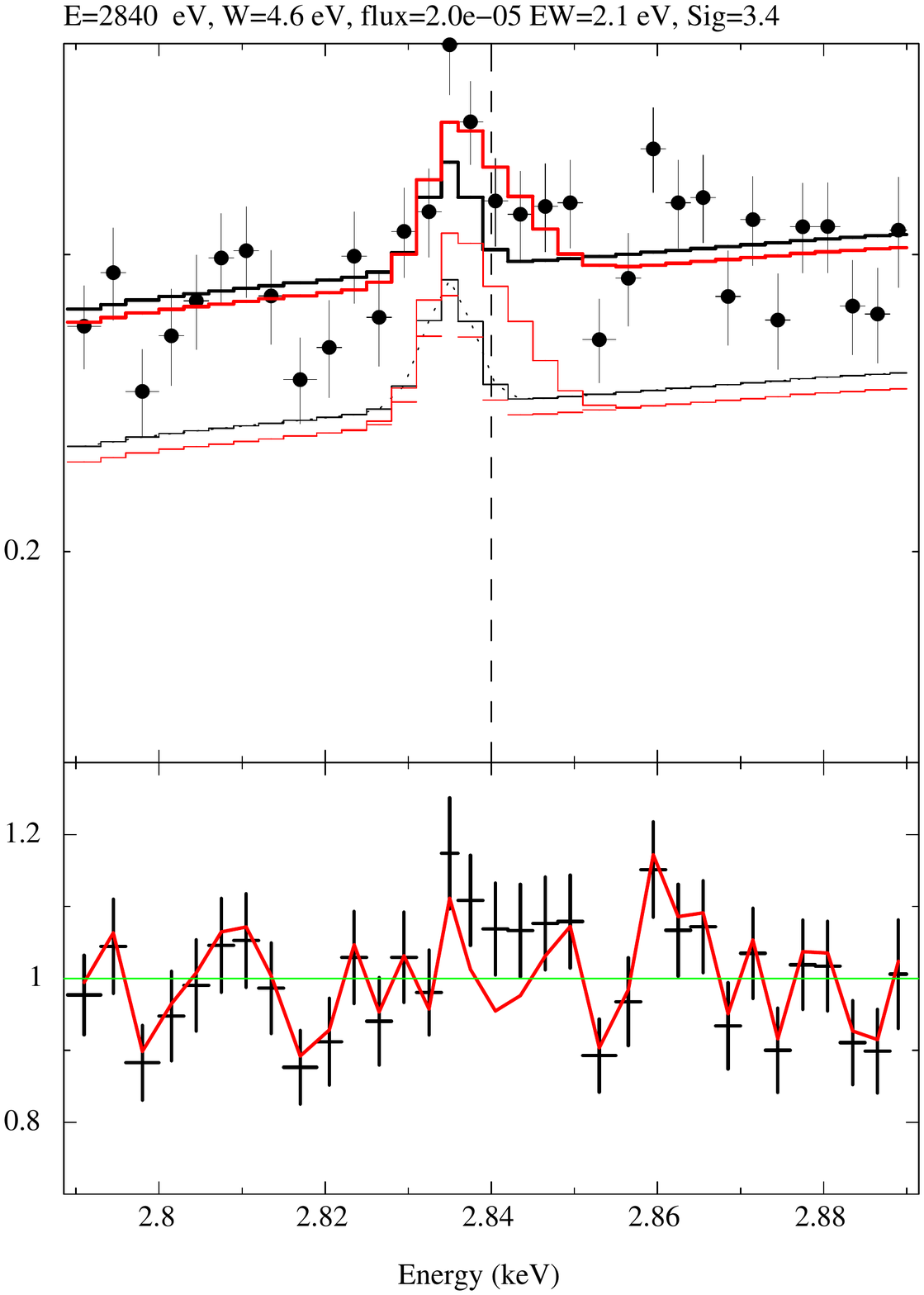}
      \includegraphics[width=0.330\textwidth]{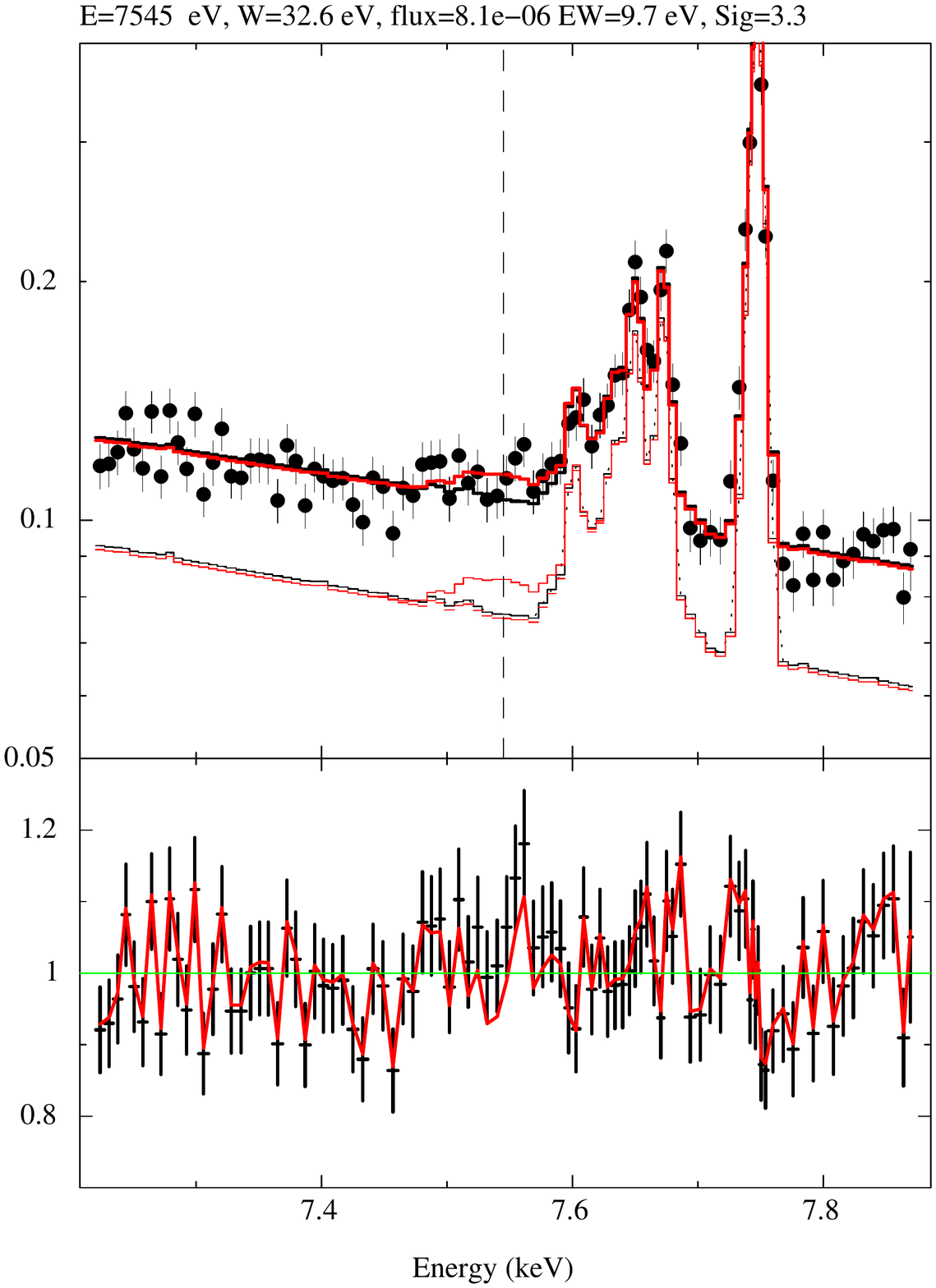}        
  }}
    \centerline{\hbox{
      \includegraphics[width=0.330\textwidth]{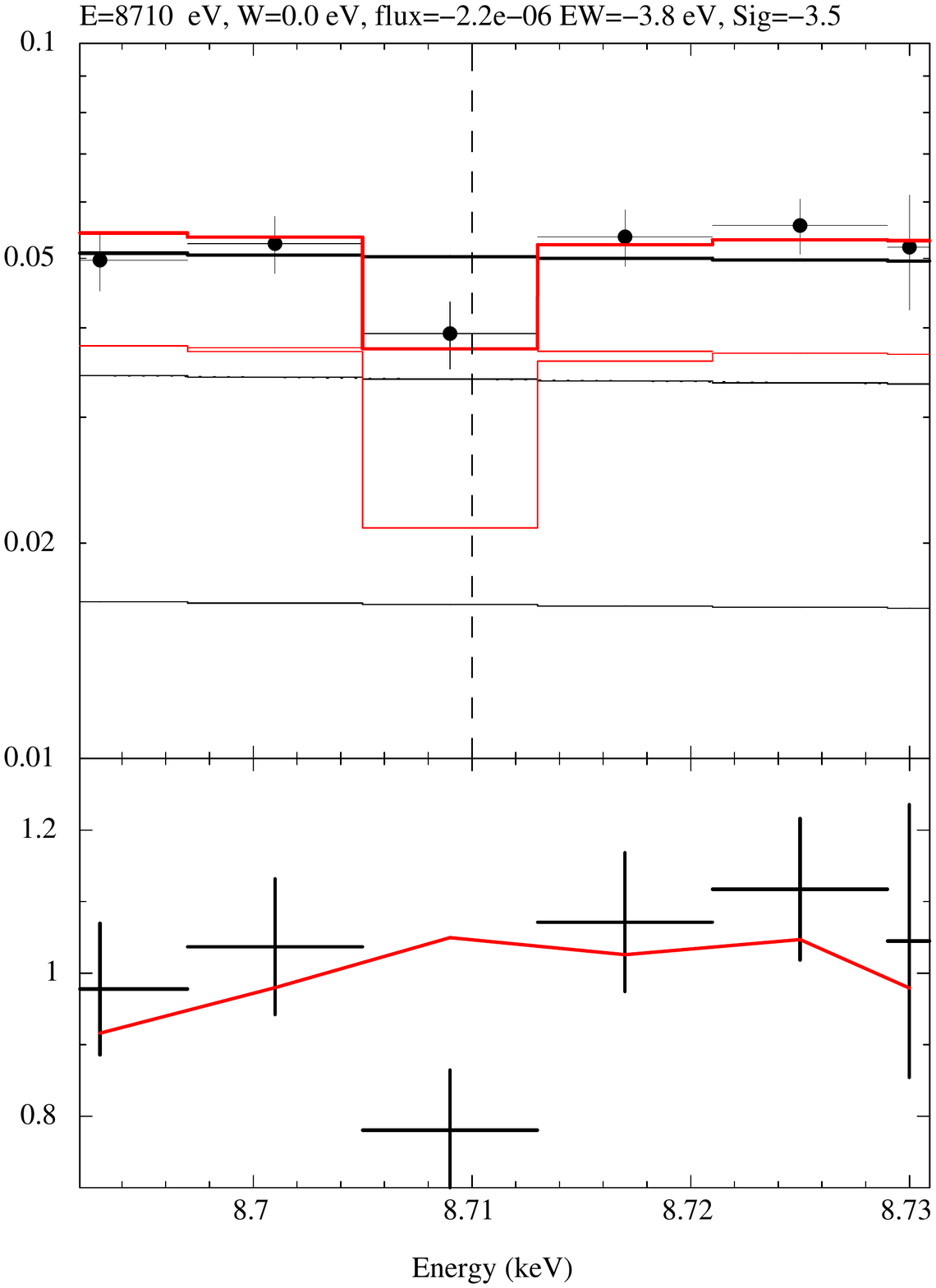}
      \includegraphics[width=0.330\textwidth]{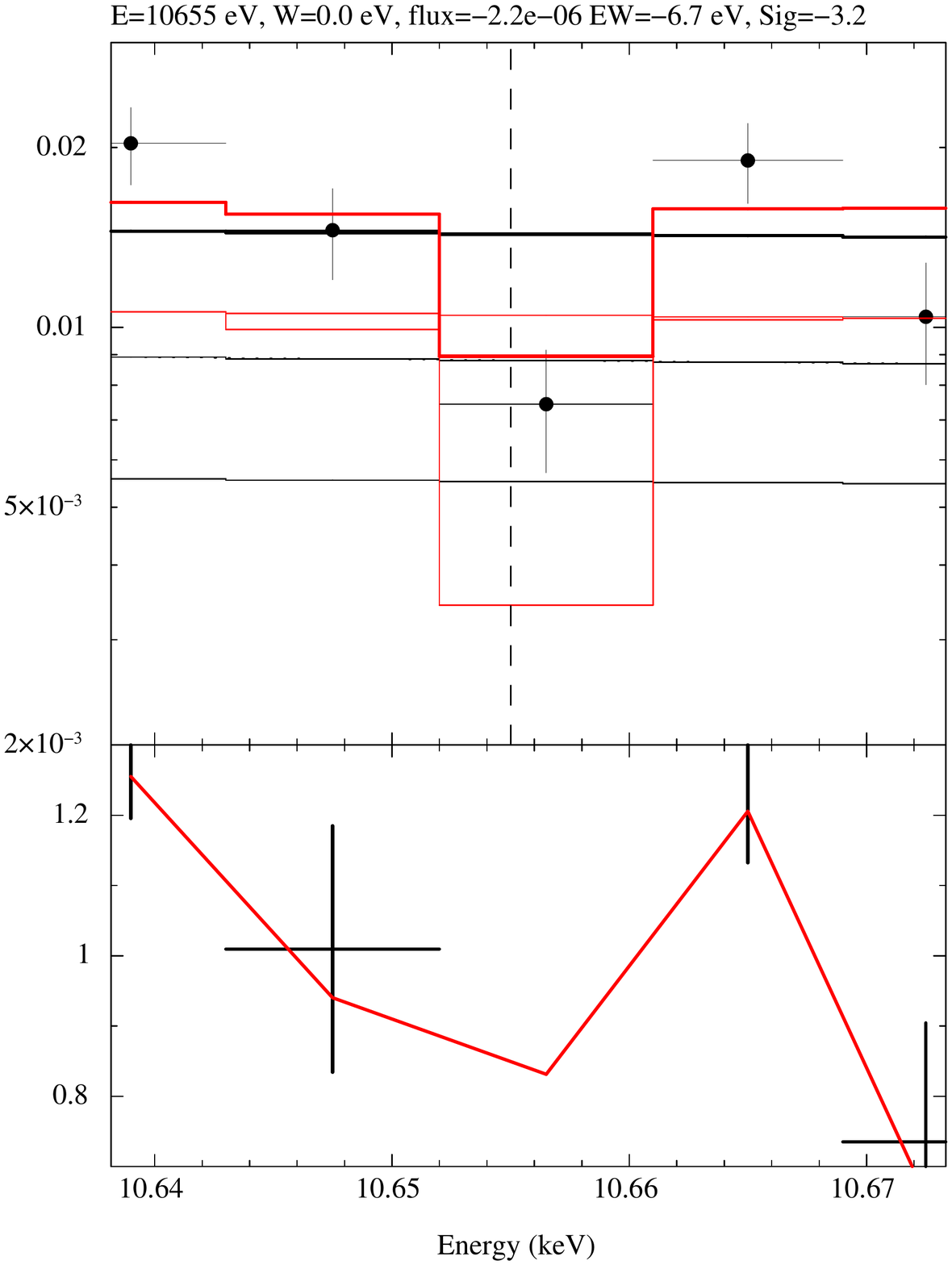}
      \includegraphics[width=0.330\textwidth]{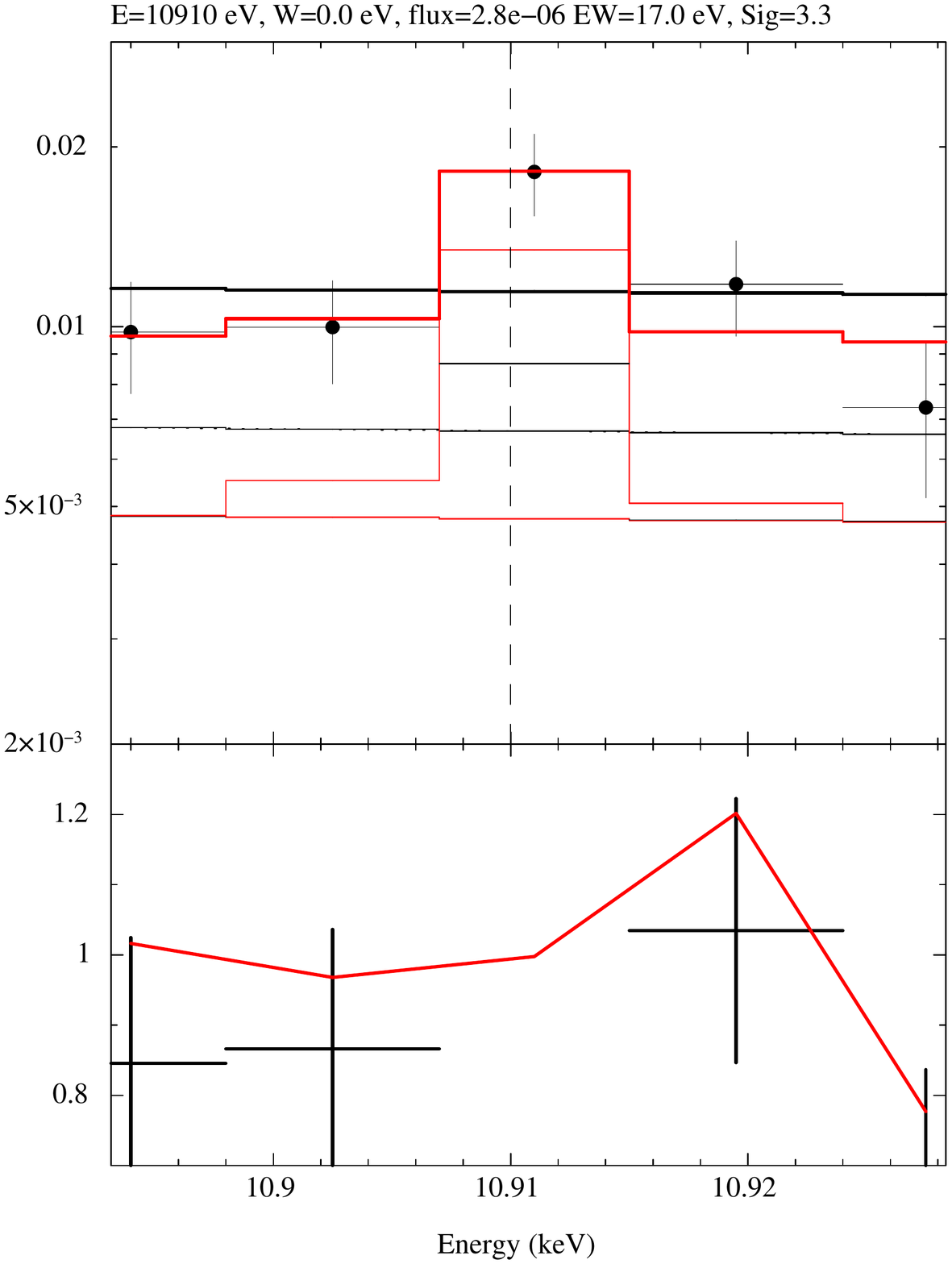}
  }}
\caption{
Zooming-in spectral fitting results
from the energy ranges where possible line emission features 
are found.
Units of top and bottom panels are
counts s$^{-1}$ keV$^{-1}$ 
and
data to model ratio, respectively.
Position and other best-fit parameters of the input Gaussian line
are given in each panel.
Black (red) histograms and residuals
are for models without (with) the line feature 
at the position shown in the dashed vertical line.
Note that spectrum is rebined coarsely for plotting purpose
but for fitting we use the original 1~eV bin size.
}\label{sf-vv:1}
\end{figure*}

\begin{figure*}[h]
  \centerline{\hbox{
      \includegraphics[width=0.33\textwidth]{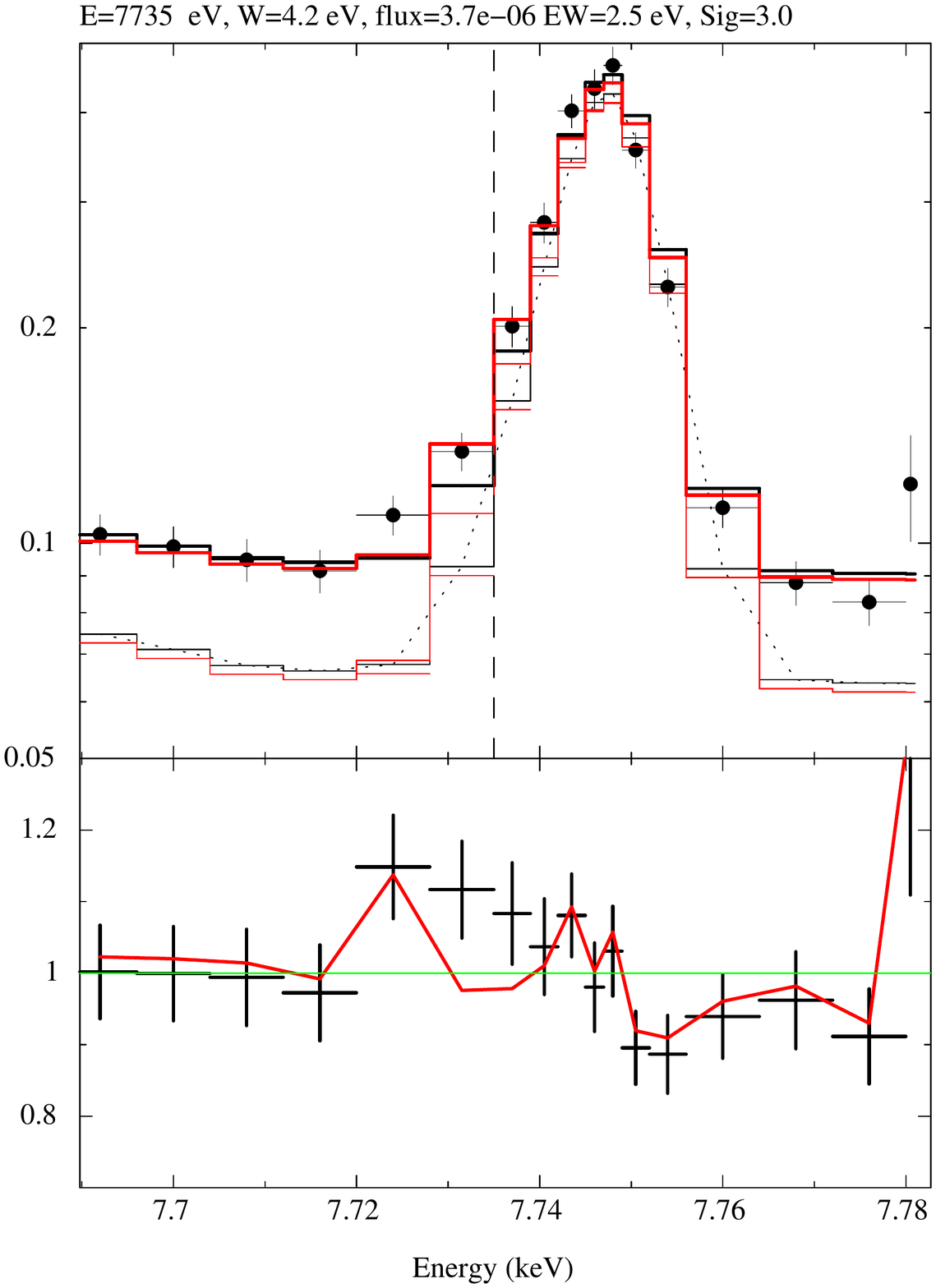}
      \includegraphics[width=0.33\textwidth]{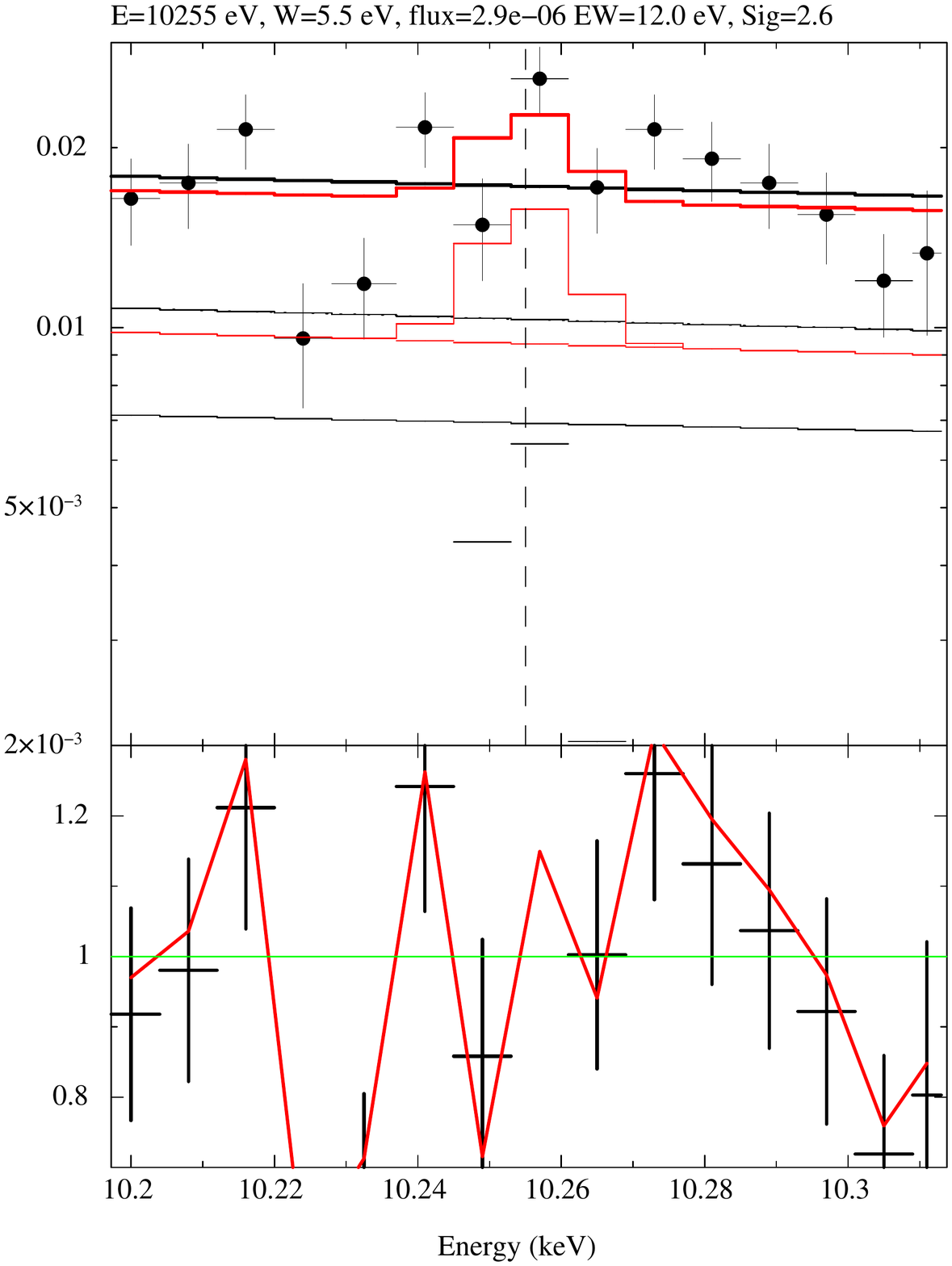}            
      \includegraphics[width=0.33\textwidth]{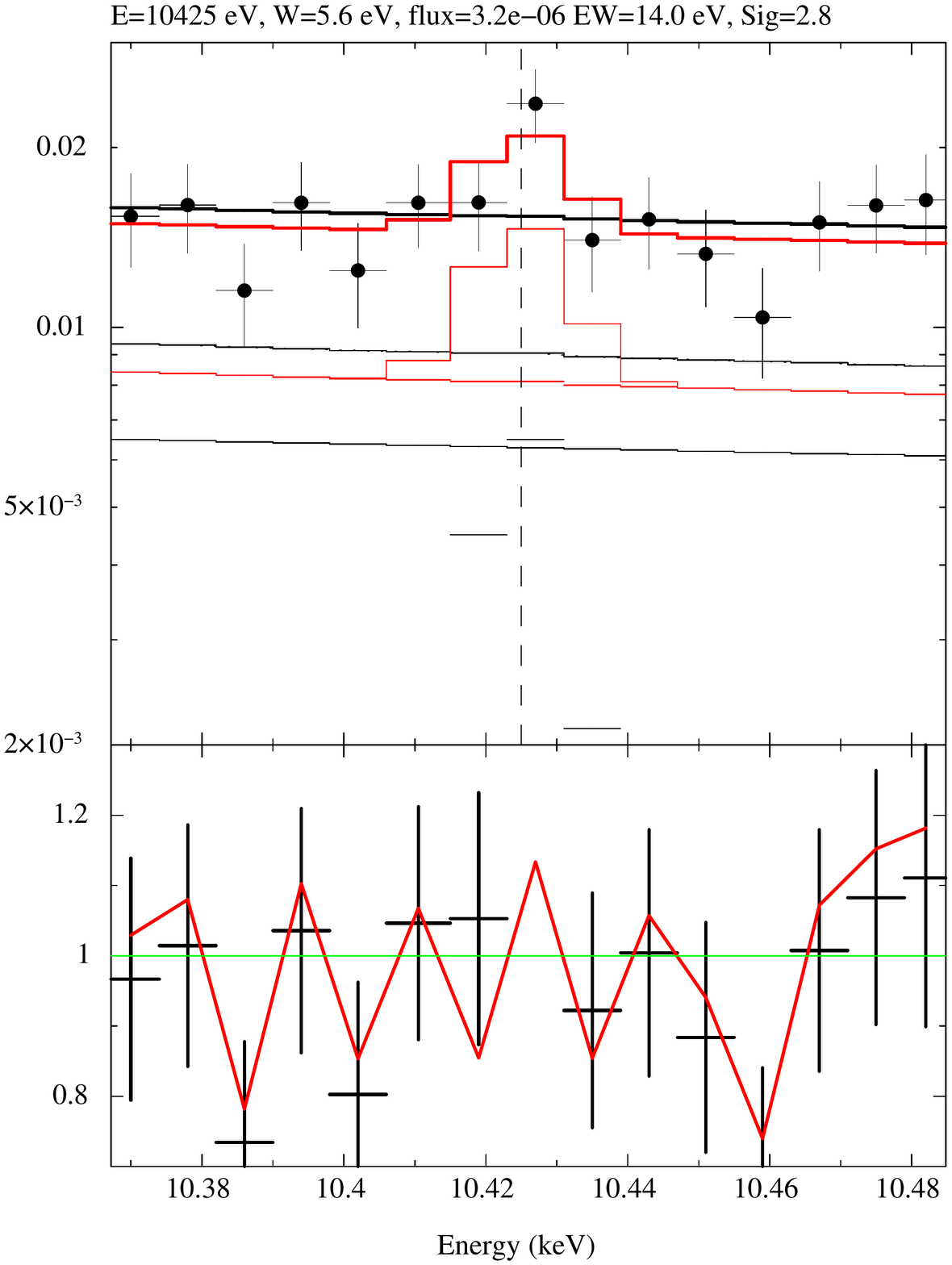}                    
  }}
\caption{
Same plots as figure~\ref{sf-vv:1}, 
but for \lgauss = 160 \kms and emission signals.
}\label{sf-v160:e}
\end{figure*}

\begin{figure*}[h]
  \centerline{\hbox{
      \includegraphics[width=0.33\textwidth]{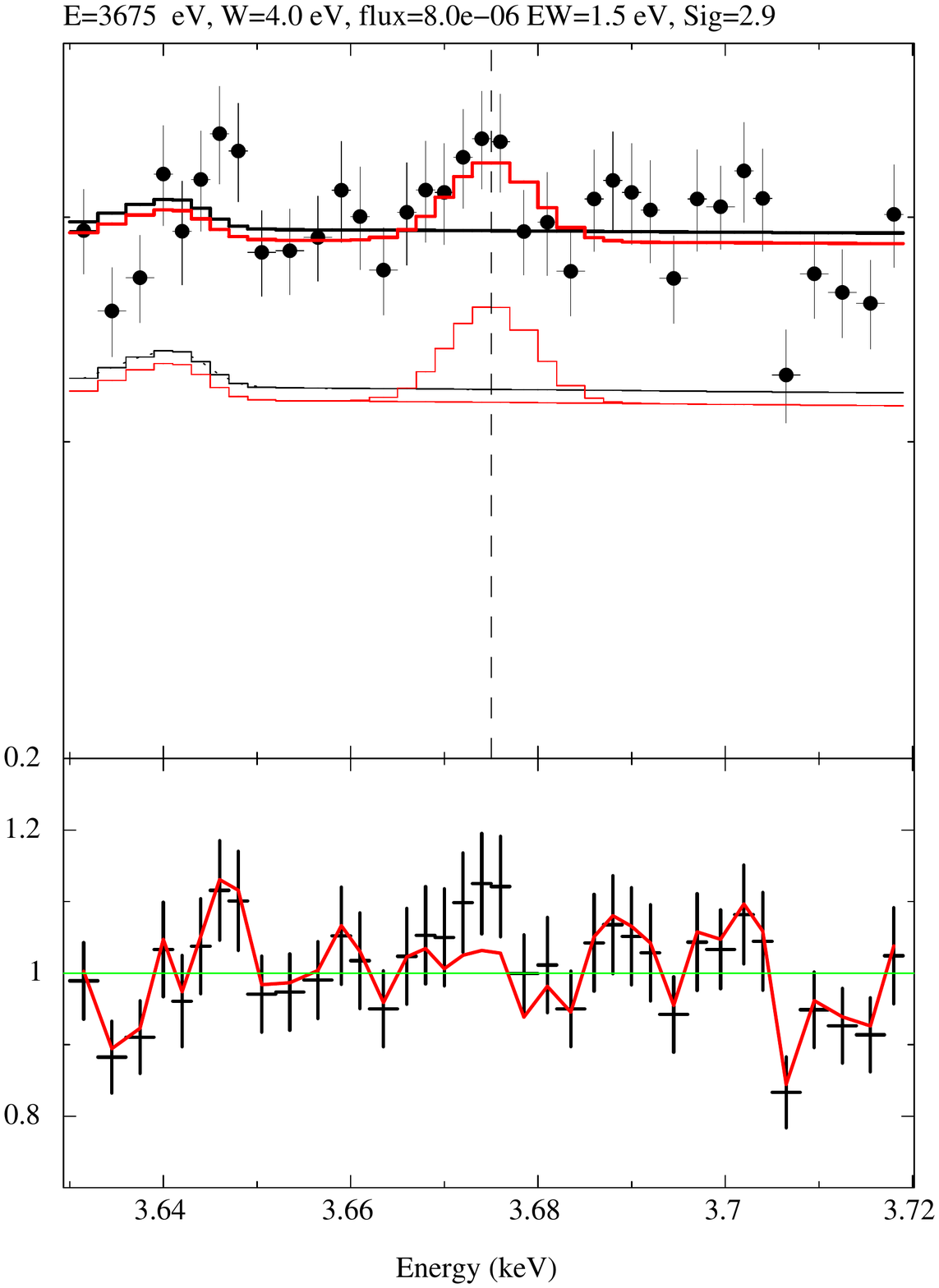}
      \includegraphics[width=0.33\textwidth]{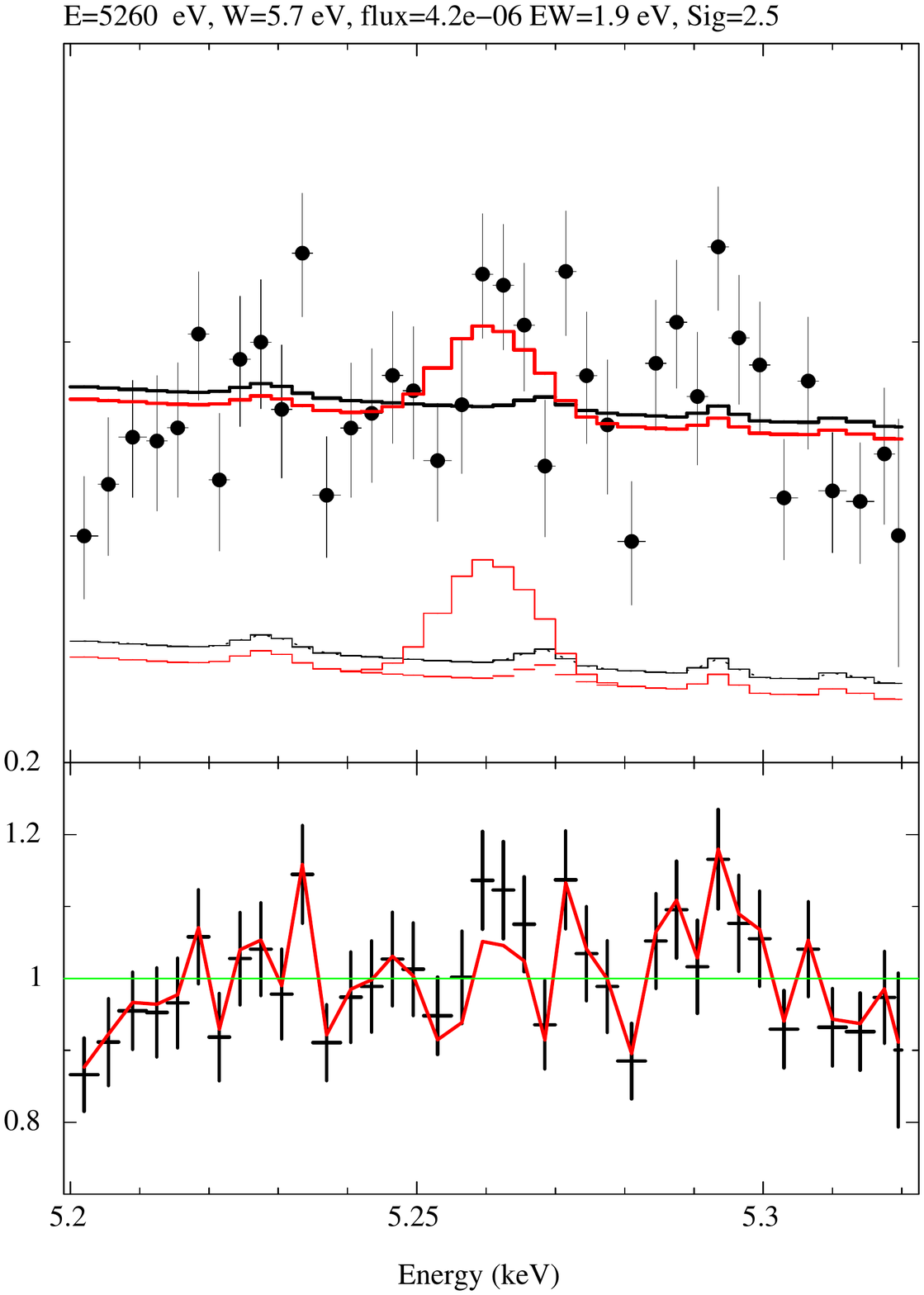}                          
      \includegraphics[width=0.33\textwidth]{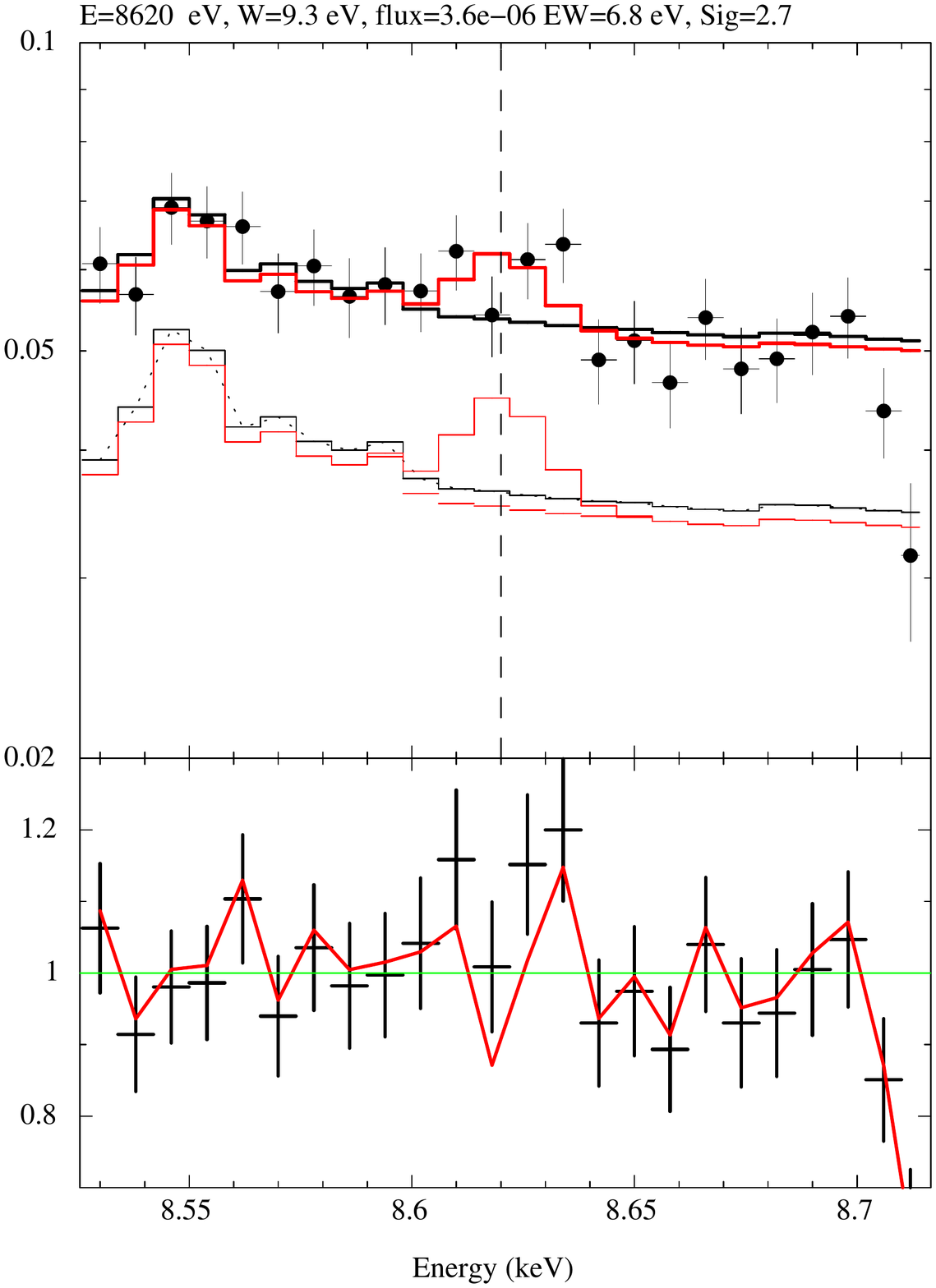}                                
  }}
\caption{
Same plots as the previous one, 
but for \lgauss = 320 \kms and emission signals.
See Table~\ref{line-tbl:1}.
}\label{sf-v160:320}
\end{figure*}

\begin{figure*}[h]
  \centerline{\hbox{
      \includegraphics[width=0.33\textwidth]{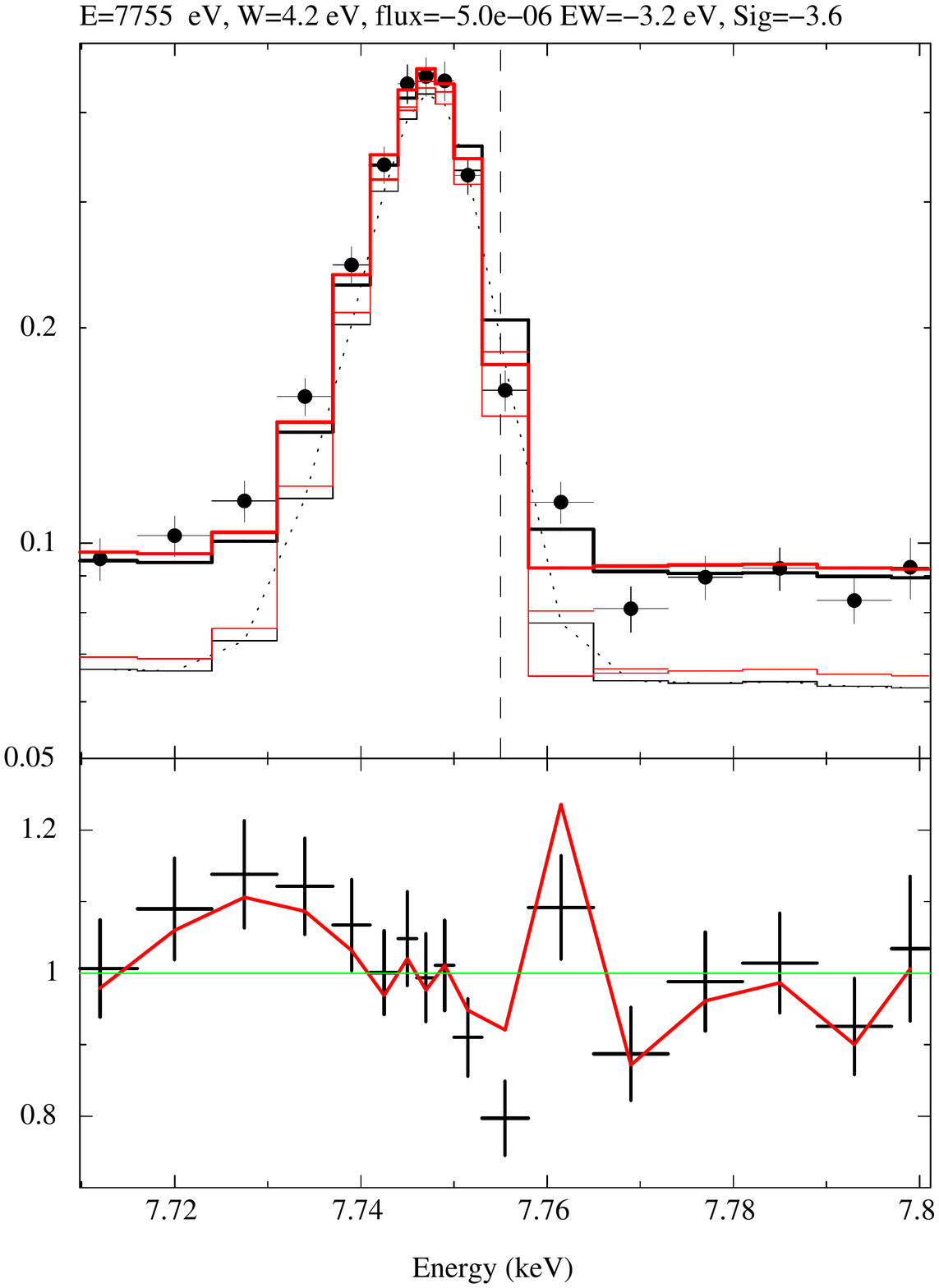}
      \includegraphics[width=0.33\textwidth]{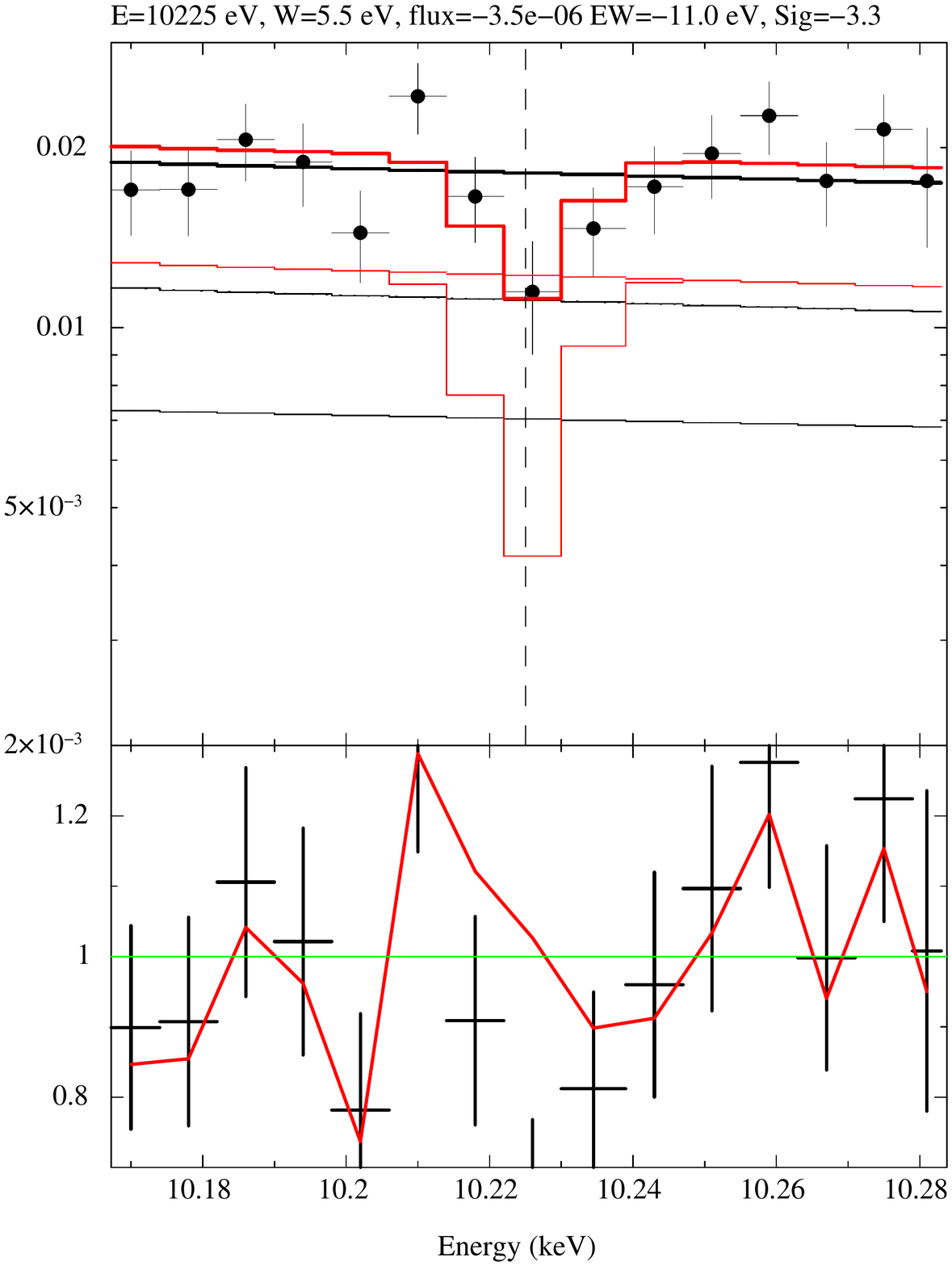}      
  }}
\caption{
Same plots as the previous one, 
but for \lgauss = 160 \kms and absorption signals.
}\label{sf-v160:320b}
\end{figure*}